\newcommand{\non}{\nonumber}
\newtheorem{theorem}{Theorem}

\def\T{\tiny\mbox{\rm T}}

\def\bfe{\bm e}

\def\bfu{\bm u}
\def\bfv{\bm v}

\def\bfy{\bm y}

\def\bfA{\bm A}
\def\bfB{\bm B}

\def\bfI{\bm I}

\def\bfK{\bm K}
\def\bfL{\bm L}
\def\bfM{\bm M}

\def\bfP{\bm P}
\def\bfQ{\bm Q}
\def\bfR{\bm R}
\def\bfS{\bm S}
\def\bfT{\bm T}
\def\bfU{\bm U}
\def\bfV{\bm V}
\def\bfW{\bm W}
\def\bfX{\bm X}
\def\bfY{\bm Y}
\def\bfZ{\bm Z}

\def\bfPhi{\mbox{\boldmath$\Phi$}}

\def\bftheta{\mbox{\boldmath${\theta}$}}

\documentclass[twocolumn]{autart}

\usepackage{algorithmic}
\usepackage{algorithm}
\usepackage{graphicx}
\usepackage{amsmath,amssymb,amsfonts}
\usepackage{MnSymbol}
\usepackage{textcomp}
\usepackage{bm}
\usepackage{color}
\usepackage{theorem}
\usepackage{multirow}

\usepackage{makecell}
\newtheorem{Remark}[theorem]{Remark}
\newtheorem{Definition}[theorem]{Definition}
 \newtheorem{Proposition}[theorem]{Proposition}
  \newtheorem{Theorem}[theorem]{Theorem}
  \newtheorem{Lemma}[theorem]{Lemma}
 \newcommand{\tabincell}[2]{\begin{tabular}{@{}#1@{}}#2\end{tabular}}
\usepackage{appendix}

\begin{document}

\begin{frontmatter}

\title{Sparse plus low-rank identification for dynamical latent-variable graphical AR models
} 

\thanks[renewcommand]{
  This work was supported by the National Natural Science Foundation of China (Grant No. 61991414, 62088101).
}
\thanks[footnote]{Corresponding author}
\author[Paestum]{Junyao You}\ead{yjy804521297@163.com},
\author[Paestum,Rome,footnote]{Chengpu Yu}\ead{yuchengpu@bit.edu.cn}.
\address[Paestum]{School of Automation, Beijing Institute of Technology, Beijing 100081, PR China}  
\address[Rome]{Chongqing Innovation Center, Beijing Institute of Technology,
Chongqing 401147, PR China}

\begin{abstract}                          
This paper focuses on the identification of
graphical autoregressive models with dynamical latent variables.
The dynamical structure of latent variables is described by a matrix polynomial transfer function.
Taking account of the sparse interactions between the observed variables
and the low-rank property of the latent-variable model,
a new sparse plus low-rank optimization problem is formulated to identify the graphical auto-regressive part, which is then handled using the trace approximation and reweighted nuclear norm minimization.
Afterwards, the dynamics
of latent variables are recovered from low-rank spectral decomposition using the trace norm convex programming method.
Simulation examples are used to illustrate the effectiveness of the proposed approach.
\end{abstract}

\end{frontmatter}

\section{Introduction}
\label{sec:introduction}
Graphical models for Gaussian stochastic processes provide a convenient visualization and inference tool to expose relative conditional independences between random variables,
and have found their applications
in statistical physics, computational biology, speech processing, statistical image processing, finance and many other fields \cite{Dahlhaus2000_MTS,Wainwright2019_HDS}.
This graphical representation allows to detect the topological
structure of the Gaussian Markov networks
by solving the covariance (extension) selection problem,
and thereby has been applied for the identification of static Gaussian models
\cite{Dempster1972_CS,Friedman2008_Glasso},
dynamical graphical
autoregressive (AR) processes \cite{Songsiri2010_TSAR,Zorzi2019_BayesianAR},
reciprocal processes
\cite{Carli2011_cerp,Lindquist2013_crce,Alpago2018_IGM}
and autoregressive moving-average (ARMA) processes \cite{Avventi2013_ARMAgraphical,Alpago2020_LP,You2022_GME_ARMA}.
When the data to be modeled by the Gaussian stochastic process has an extremely high dimension,
the corresponding graphical model could be complicated with most of
the components genuinely interconnected
and can only
provide limited information of the model structure
\cite{Alpago2017_SPGM}.
This may \textcolor{blue}{be} due to
the presence of latent variables that are not observed but
cause a common behavior in all the observed variables \cite{Chandrasekaran2012_lmco}.
Therefore,
by representing such graphical models as latent-variable
graphical models,
we may infer the complete network structure
according to the extra hidden structure information contained in latent-variable dynamics.

 Under the assumption that the observed variables become nearly independent when conditioned on the latent variables,
latent-variable graphical models admit a sparse plus low-rank structure with the sparse structure accounting for few direct interactions between the observed variables
and the low-rank structure modeling the effect of latent variables.
The detection of the model structure is thereby transformed into a sparse plus low-rank decomposition problem.
Basic sparse plus low-rank decomposition problem
can be formulated as a convex program
by minimizing a linear combination of the $l_1$ norm and the
nuclear norm of the components \cite{Wright2022_HD},
which can be solved by the convex programming methods.
Based on this paradigm,
many methods have been developed for the identification
of latent-variable graphical models.
For example,
a regularized maximum entropy based method was proposed to estimate conditional dependency structure of latent-variable graphical models associated with Gaussian AR processes by exploiting the
sparse plus low-rank decomposition of the
inverse spectral density matrix corresponding to the observed variables \cite{Zorzi2016_ARlatent,Liegeois2015_Sparse_low_rank}.
In this identification scheme,
two regularizers are considered to separately induce a small number of latent nodes and sparsity in the network.
The numerical complexity of
high-order AR processes in latent-variable graphical models
was addressed in \cite{Alpago2022_lgraphical}
by using reciprocal approximation.
In \cite{Ciccone2020_ARlatent},
a new sparse plus low-rank identification paradigm was established,
where the uncertainty in the estimation
was depicted by a confidence neighborhood containing the true model computed with a prescribed probability.
The resulting optimization model involves just one regularization parameter balancing the tradeoff between the sparseness of the learned
graphical structure and the number of latent variables.
The cross-validation procedure for choosing the
optimal regularization parameter could be simplified with respect to the method proposed in \cite{Zorzi2016_ARlatent},
thus reducing the computational burden.

Among those aforementioned works, \cite{Zorzi2016_ARlatent,Liegeois2015_Sparse_low_rank,Alpago2022_lgraphical,Ciccone2020_ARlatent}
construct time-domain optimization models (formulated in matricial form) that are amenable to variational analysis to learn the graphical AR network structure
in the presence of latent nodes.
To deal with the sparse plus low-rank decomposition of
the inverse spectrum,
\cite{Veedu2022_LDM} provided a frequency-domain method for
reconstructing the topology of networked linear dynamical systems with latent nodes.
Necessary and sufficient
conditions for unique sparse plus low-rank decomposition
of a skew symmetric matrix corresponding to the imaginary part of the inverse spectrum were given based on the rank-sparsity incoherence notion,
which extends the results in \cite{Chandrasekaran201_incoherence}.
The identified sparse component could yield the moral
graph formed by the observed nodes,
and the low-rank component could retrieve the Markov Blanket of the hidden nodes.
It has been claimed that the frequency-domain approach proposed in \cite{Veedu2022_LDM}
can be employed towards the retrieval of the
moral graph of networks of AR models and has been applied to
networked systems \cite{Zorzi2016_ARlatent,Ciccone2020_ARlatent}.
Further research \cite{Veedu2020_spatial}
was carried out to generalize this method to the case of exogenous excitements
with hidden nodes and spatially correlated noise.

\textcolor{blue}{In this paper,
we focus on the joint estimation of parameters and graph topology of graphical AR models
with dynamic latent variables.
It is stipulated that most of the interconnections among observable nodes in a graphical model are generated by a small number of dynamic latent variables.
The dynamic structure of latent variables is inspired from the moving-average (MA) factor models \cite{Falconi2021_MA_factor,Falconi2021_ARMA_factor},
which can organize the available data into a concise and parsimonious structured representation.
Therefore, the latent-variable graphical model can provide a better interpretation for complex high-dimensional data.
For the concerned identification problem,
the direct implementation of the sparse plus low-rank decomposition framework in  \cite{Zorzi2016_ARlatent,Ciccone2020_ARlatent} is infeasible
due to the failure of matricial reformulation for the low-rank structure.
Although the methods proposed in \cite{Veedu2022_LDM,Veedu2020_spatial}
could deal with the topology reconstruction problem for the concerned problem,
they did not perform the parameter identification.}
The main contributions of this paper are listed as follows.
\begin{itemize}
\item
 A new sparse plus low-rank optimization framework is proposed by exploiting the sparse structure 
 characterizing the interdependence among the observed variables and the low-rank property for the power spectral expression of the latent variables,
 which can reveal the conditional graphical structure formed by observed variables and yield the AR parameter estimates.

 \item
After obtaining the topology estimate, the associated low-rank optimization problem is reformulated using the Shur complement lemma, for which finer results are obtained by adopting an iteratively reweighted trace minimization approach.

 \item
 A low-rank spectral factorization problem is formulated to learn the dynamic latent-variable model by taking into account of the estimation uncertainty described by an empirically computed tolerance.
 By deriving its matricial reparametrization,
 we are able to recover the number of latent variables and the corresponding dynamics.
\end{itemize}

The rest of the paper is organized as follows.
Section~\ref{sec:Notation} gives the notation and preliminaries used throughout the paper.
Section~\ref{sec:Formulation} introduces graphical AR models with dynamical latent variables and formulates the identification problem.
Section~\ref{sec:Results} presents a sparse plus low-rank identification algorithm for latent-variable graphical AR models.
Section~\ref{sec:Simulation} provides two simulation examples to demonstrate the effectiveness of the proposed algorithm.
Finally, conclusions are drawn in Section~\ref{sec:Conclusions}.
\section{Notation and Preliminaries}
\label{sec:Notation}
In this paper,
${\boldsymbol{I}_n}$ represents the $n\times n$ identity matrix and
${\boldsymbol{0}}_{m\times n}$ denotes the set of $m\times n$ matrices whose entries are all zeros.
For a matrix $\bfX$,
$\bfX^{\T}$ or $[\bfX]^{\T}$ represents its transpose,
${\rm{rank}}(\bfX)$ denotes its rank,
and $[\bfX]_{ij}$ or $\bfX_{ij}$ represents its $ij$-th entry.
${\rm{supp}}(\bfX)$ is the support set defined as $\{(i,j):\bfX_{ij}\neq 0\}$,
 and the number of non-zero entries in $\bfX$ is denoted by $\vert{\rm{supp}}(\bfX)\vert$ or $\|\bfX\|_0$.
If $\bfX$ is a square matrix,
${\rm{tr}}(\bfX)$ denotes its trace,
 and
 ${\rm{diag}}({\boldsymbol{X}})$ represents a diagonal matrix with elements on its main diagonal.
 If matrix $\bfX$ is positive definite (semi-definite), we write
 $\bfX> 0$ ($\bfX\geq 0$).
For square matrices $\bfA,\bfB$ with the same dimension, we define their inner product as $\langle\bfA,\bfB\rangle:={\rm{tr}}(\bfA^{\T}\bfB)$.
The symbols $\|\cdot\|_F$ and $\|\cdot\|_\ast$ stand for the Frobenius norm and nuclear norm, respectively.
\textcolor{blue}{The symbol ${\rm{Pr}}(A)$ denotes the probability of the event $A$.}

Symbols $\mathbb{R}^{m\times n}$ and $\mathbb{R}^{n}$ denote the set of real matrices of size $m\times n$ and real column vectors of size $n\times 1$, respectively.
$\mathbb{S}^{n}$ is the set of real symmetric matrices of order $n$.
\textcolor{blue}{
$\mathbb{M}^{n,p}$ is the set of matrices $\bfQ:=[\bfQ_0\ \bfQ_1\ \cdots\ \bfQ_p]$ with
$\bfQ_0\in\mathbb{S}^{n}$ and $\bfQ_j\in\mathbb{R}^{n\times n}$ $(j=1,\cdots,p)$.
The linear mapping $\mathcal{T}$ from $\mathbb{M}^{n,p}$ to $\mathbb{S}^{n(p+1)}$ constructs a symmetric block Toeplitz matrix from its first block row:
if $\bfQ\in\mathbb{M}^{n,p}$, then
\begin{eqnarray}
\mathcal{T}(\bfQ)=\setlength\arraycolsep{1pt}
        \left[\begin{array}{cccc}
           \bfQ_{0}     &  \bfQ_{1}        & \cdots & \bfQ_{p} \\
             \bfQ_{1}^{\T}        &  \bfQ_{0}       &\cdots  & \bfQ_{p-1}\\
             \vdots    &\vdots    & \ddots & \vdots\\
            \bfQ_{p}^{\T}         & \bfQ_{p-1}^{\T}    & \cdots      & \bfQ_{0} \end{array}\right].\non
\end{eqnarray}
The adjoint of $\mathcal{T}$ is a mapping $\mathcal{D}$ from $\mathbb{S}^{n(p+1)}$ to $\mathbb{M}^{n,p}$ constructing a vector space:
if the matrix $\bfS\in\mathbb{S}^{n(p+1)}$ is partitioned as
\begin{eqnarray}
\label{bfX_partition}
\bfS=\setlength\arraycolsep{1pt}
        \left[\begin{array}{cccc}
           \bfS_{0,0}     &  \bfS_{0,1}        & \cdots & \bfS_{0,p} \\
             \bfS_{0,1}^{\T}        &  \bfS_{1,1}       &\cdots  & \bfS_{1,p}\\
             \vdots    &\vdots    & \ddots & \vdots\\
            \bfS_{0,p}^{\T}         & \bfS_{1,p}^{\T}    & \cdots      & \bfS_{p,p} \end{array}\right],
\end{eqnarray}
where $\bfS_{i,i}\in\mathbb{S}^{n}$ and
$\bfS_{i,j}\in\mathbb{R}^{n\times n}$ $(i\neq j)$ denote the sub-blocks of $\bfS$,
then
\begin{eqnarray}
\mathcal{D}(\bfS)&:=&[\mathcal{D}(\bfS)_0\ \mathcal{D}(\bfS)_1\ \cdots\ \mathcal{D}(\bfS)_p]\in \mathbb{M}^{n,p},\non\\
\mathcal{D}(\bfS)_0&=&\sum_{v=0}^{p}\bfS_{v,v},\ \mathcal{D}(\bfS)_j=2\sum_{v=0}^{p-j}\bfS_{v,v+j},\ j=1,\cdots,p.\non
\end{eqnarray}}

Let $\mathbb{E}$ denote the expectation operator.
$F^\ast(z)=F(z^{-1})^{\T}$ represents the \textcolor{blue}{Hermitian transpose.
Define the family ${\mathcal Y}(n,p)$ of $n\times n$ matrix pseudo-polynomials of order $p$ as
\begin{eqnarray}
 {\mathcal Y}(n,p)&=&\bigg\{Y(z)={\mathcal R}\bigg\{\sum_{j=0}^{p}z^{-j}\bfY_j \bigg\}:\ \bfY_j\in \mathbb{R}^{n\times n},\non\\
 &&\ Y(e^{{\rm{j}} \omega})>0, \ \forall \omega\in[-\pi,\pi] \bigg\},\non\\
{\mathcal R}\{D(z)\}&:=&\frac{1}{2}[D(z)+D^\ast(z)].\non
\end{eqnarray}}
Note that we take the operator $z=e^{{\rm{j}} \omega}$ in this paper and the dependence upon $z$ will be dropped if not needed.

Let $\mathcal G(\mathcal V,\mathcal E)$ denote a graph,
where $\mathcal V=\{1,\cdots,n\}$ is the set of vertices and $\mathcal E\subseteq \mathcal V\times \mathcal V$ is the set of edges.
For a zero-mean stationary Gaussian process $\boldsymbol{z}:=\{{\boldsymbol{z}}(t)\in\mathbb{R}^{n+l},\ t\in \mathbb{Z}\}$ with $n$
observable variables and $l$ latent variables,
the corresponding latent-variable graphical model is denoted by $\mathcal G(\mathcal V_{n+l},\mathcal E_{n+l})$.
Specifically,
the vertices represent the variables of the process  $\boldsymbol{z}:=[{\boldsymbol{y}}^{\T},{\boldsymbol{x}}^{\T}]^{\T}
\ ({\boldsymbol{y}}\in\mathbb{R}^{n},\ {\boldsymbol{x}}\in\mathbb{R}^{l})$,
and the edges describe the conditional dependence relations among the variables through
\begin{eqnarray}
(k,q)\notin\mathcal E_{n+l} \Leftrightarrow \mathcal X_{\{k\}}\upmodels \mathcal X_{\{q\}}|\mathcal X_{\{\mathcal V_{n+l}\backslash\{k,q\}\}},\non
\end{eqnarray}
where ${\mathcal X}_{\mathcal I}:={\rm{span}}\{{\boldsymbol{z}}_k(t): k\in \mathcal I; \ t\in \mathbb{Z}\}$ for $\mathcal I \subset \mathcal V_{n+l}$.
That is,
the absence of
the edge between node $k$ and $q$ implies that
for all $t_1$, $t_2$,
the variables ${\boldsymbol{z}}_k(t_1)$ and ${\boldsymbol{z}}_q(t_2)$
are conditionally independent given the space spanned by $\{{\boldsymbol{z}}_i(t): i\in \mathcal V_{n+l}\backslash\{k,q\}\}$.
We denote the number of conditional dependence pairs among observable variables by  $\vert\mathcal E_n\vert$.

\section{Problem Formulation}
\label{sec:Formulation}
Consider the following multivariate graphical \textcolor{blue}{AR Gaussian processes} with latent variables:
\begin{eqnarray}
\label{AR_dyl}
{\bfA}(z){\boldsymbol{y}}(t)&=&{\boldsymbol{W}}_L(z){\boldsymbol{x}}(t)+{\boldsymbol{\omega}}(t),\\
{\boldsymbol{A}}(z)&:=&{\boldsymbol{I}_n}+{\boldsymbol{A}}_1z^{-1}+{\boldsymbol{A}}_2z^{-2}+\cdots+{\boldsymbol{A}}_{p_1}z^{-p_1},\non\\
{\boldsymbol{W}}_L(z)&:=&{\boldsymbol{W}}_{L,0}+{\boldsymbol{W}}_{L,1}z^{-1}+\cdots+{\boldsymbol{W}}_{L,p_2}z^{-p_2},\non
\end{eqnarray}
where ${\boldsymbol{y}}(t)\in\mathbb{R}^{n}$ is the output of the system,
${\boldsymbol{x}}(t)\in\mathbb{R}^{l}$ and ${\boldsymbol{w}}(t)\in\mathbb{R}^{n}$
are independent \textcolor{blue}{white Gaussian random vectors} with spectral density matrices ${\bf\Phi}_{x}(z)={\boldsymbol{I}}_l$ and
 ${\bf\Phi}_{w}(z)={\boldsymbol{I}}_n$. 
${\boldsymbol{A}}_j\in\mathbb{R}^{n\times n}\ (j=1,2,\cdots,p_1)$
are sparse parameter matrices of the graphical AR processes,
and
${\boldsymbol{W}}_{L,i}\in\mathbb{R}^{n\times l}\ (i=0,1,\cdots,p_2)$
are parameter matrices of the $l$-dimensional latent (or
hidden) variables ${\boldsymbol{x}}(t)$.
The latent-variable graphical AR models (\ref{AR_dyl}) can be seen as the extension of factor models to graphical models,
\textcolor{blue}{
where the dynamical structure of
each component of
observable ${\boldsymbol{y}}(t)$ is determined by
the sum of a component of the latent process ${\bfA}(z)^{-1}{\boldsymbol{W}}_L(z){\boldsymbol{x}}(t)$ driven by $l$ common factors ${\boldsymbol{x}}(t)$
and a component of idiosyncratic noise ${\bfA}(z)^{-1}{\boldsymbol{w}}(t)$
\cite{Falconi2021_MA_factor,Falconi2021_ARMA_factor,Pan2008_ModellingMT,Ciccone2019_FMR,Crescente2020_AR_factor}.}
Note that both ${\boldsymbol{x}}(t)$ and ${\boldsymbol{w}}(t)$ are unobservable.
In this paper, it is stipulated that $l\leq n$, the orders $p_1$ and $p_2$ are known,
${\boldsymbol{y}}(t)=0$, ${\boldsymbol{x}}(t)=0$, and ${\boldsymbol{w}}(t)=0$ for $t\leqslant0$.
The identification objective is to estimate the unknown parameter matrices ${\boldsymbol{A}}_j$ and ${\boldsymbol{W}}_{L,i}$ according to  the observations ${\boldsymbol{y}}(t)$,
following the determination
of the conditional independence relations
among the observable variables ${\boldsymbol{y}}(t)$ coded in graphical AR processes as well as the number of latent variables ${\boldsymbol{x}}(t)$.
\begin{Remark}
\label{Remark1}
Different from the latent-variable graphical models studied in \cite{Zorzi2016_ARlatent,Ciccone2020_ARlatent},
we model the dynamical effect of the latent variables by a matrix polynomial ${\boldsymbol{W}}_L(z)$,
which admits a more concrete structure explaining the interconnections among observable variables.
This dynamic latent-variable model makes it difficult to derive a matricial reformulation of the sparse plus low-rank decomposition optimization paradigm.
By separating the observable variables and latent variables,
Model (\ref{AR_dyl}) can be expressed as an unilateral $z$-transform transfer function model determined by finite length sequences similar to the linear dynamical model studied in \cite{Veedu2022_LDM}.
While the work \cite{Veedu2022_LDM} focuses on the topology reconstruction problem,
we consider the joint estimation of parameters and graph topology based on sampled data.
\end{Remark}
Let the power spectral density matrix ${\bf\Phi}(z)$ of the whole process with $n$
observable variables and $l$ latent variables be decomposed of
\begin{eqnarray}
\setlength\arraycolsep{1pt}
       {\bf\Phi}(z)
=\setlength\arraycolsep{2pt}
        \left[\begin{array}{cc}
           {\bf\Phi}_y(z)  &  {\bf\Phi}_{yx}(z)   \\
             {\bf\Phi}_{xy}(z)        &  {\bf\Phi}_{x}(z)       \end{array}\right],\
  {\bf\Phi}(z)^{-1}
=\setlength\arraycolsep{2pt}
        \left[\begin{array}{cc}
           {\bf\Upsilon}_y(z)  &  {\bf\Upsilon}_{yx}(z)   \\
             {\bf\Upsilon}_{xy}(z)        &  {\bf\Upsilon}_{x}(z)       \end{array}\right].  \non
\end{eqnarray}
Using Schur's complement of matrix \cite{Fijany1995_Schur},
we have the following representation of
the spectrum ${\bf\Phi}_y(z)$ associated with the observable variables:
\begin{eqnarray}
\label{Phi_y_iSchur}
{\bf\Phi}_y(z)^{-1}={\bf\Upsilon}_y(z)-{\bf\Upsilon}_{yx}(z){\bf\Upsilon}_{x}(z)^{-1}{\bf\Upsilon}_{xy}(z).
\end{eqnarray}
From (\ref{AR_dyl}), the spectral density matrix ${\bf\Phi}_y(z)$  satisfies
\begin{eqnarray}
\begin{split}
\label{AR_dyl_PDM}
{\bfA}(z){\bf\Phi}_y(z){\bfA}^\ast(z)
=&{\boldsymbol{W}}_L(z){\bf\Phi}_{x}(z){\boldsymbol{W}}_L^\ast(z)+{\bf\Phi}_{w}(z)\\
=&{\bf\Phi}_{W_L}(z)+{\boldsymbol{I}}_n,
\end{split}
\end{eqnarray}
 where ${\bf\Phi}_{W_L}(z)={\boldsymbol{W}}_L(z){\boldsymbol{W}}_L^\ast(z)\in{\mathcal Y}(n,p_2)$.
 It is worth noting that ${\bf\Phi}_{W_L}(z)\geq 0$ and has rank $l$ for all $|z|\leq 1$.
 According to (\ref{AR_dyl_PDM}),
 the inverse spectrum ${\bf\Phi}_y(z)^{-1}$ can be expressed as
 \begin{eqnarray}
\label{AR_dyl_PDM2}
&&{\bf\Phi}_y(z)^{-1}\non\\
&&={\bfA}^\ast(z)[{\boldsymbol{W}}_L(z){\bf\Phi}_{x}(z){\boldsymbol{W}}_L^\ast(z)+{\bf\Phi}_{w}(z)]^{-1}{\bfA}(z).
\end{eqnarray}
Combing (\ref{Phi_y_iSchur}) with (\ref{AR_dyl_PDM2}) and applying
block matrix inversion formula \cite{Horn2012_Matrix},
we obtain a sparse plus low-rank structure for ${\bf\Phi}_y(z)^{-1}$:
\begin{eqnarray}
\label{Phi_SPhi_L}
{\bf\Phi}_y(z)^{-1}&=&{\bf\Phi}_S(z)-{\bf\Phi}_L(z),\\
\label{Phi_S}
{\bf\Phi}_S(z)&=&{\boldsymbol{A}}^\ast(z){\bf\Phi}_{w}(z)^{-1}{\boldsymbol{A}}(z)={\boldsymbol{A}}^\ast(z){\boldsymbol{A}}(z),\\
\label{Phi_L}
{\bf\Phi}_L(z)&=&{\boldsymbol{\Psi}}^\ast(z){\boldsymbol{\Lambda}}(z)^{-1}{\boldsymbol{\Psi}}(z),\\
\label{Psi}
{\boldsymbol{\Psi}}(z)&=&{\boldsymbol{W}}_L^\ast(z){\bf\Phi}_{w}(z)^{-1}{\boldsymbol{A}}(z)={\boldsymbol{W}}_L^\ast(z){\boldsymbol{A}}(z),\\
\label{Lambda}
{\boldsymbol{\Lambda}}(z)&=&{\bf\Phi}_{x}(z)^{-1}+{\boldsymbol{W}}_L^\ast(z){\bf\Phi}_{w}(z)^{-1}{\boldsymbol{W}}_L(z)\non\\
&=&{\boldsymbol{I}}_l+{\boldsymbol{W}}_L^\ast(z){\boldsymbol{W}}_L(z),
\end{eqnarray}
where ${\bf\Phi}_S$ is sparse and ${\bf\Phi}_L$ is low-rank.
\begin{Remark}
\label{Remark2}
For a multivariate
stationary \textcolor{blue}{Gaussian process} with the associated graphical structure $\mathcal G$ having full node observability,
it has
\begin{eqnarray}
\label{sparsity1}
\mathcal X_{\{k\}}\upmodels \mathcal X_{\{q\}}|\mathcal X_{\{\mathcal V\backslash\{k,q\}\}} \Leftrightarrow  [{\bf{\Phi}}(e^{{\rm{j}} \omega})^{-1}]_{kq}=0, \forall \omega \in [-\pi,\pi],\non
\end{eqnarray}
which relates the graphical structure to
the sparsity pattern of the inverse spectrum. In
the absence of the edge between node $k$ and $q$, the $kq$-th entry in the inverse spectral density matrix will be zero
 \cite{Dahlhaus2000_MTS,Avventi2013_ARMAgraphical}.
\end{Remark}
\begin{Remark}
\label{Remark3}
For the concerned multivariate
stationary \textcolor{blue}{Gaussian process} with latent variables (\ref{AR_dyl}),
the sparse part ${\bf\Phi}_S$ representation (\ref{Phi_S})
of ${\bf\Phi}_y^{-1}$
 is the same with the inverse spectrum ${\bf{\Phi}}^{-1}$ of the full variable observability case
\cite{Veedu2022_LDM,Songsiri2010_gmar}.
Therefore,
the sparsity pattern of ${\bf\Phi}_S$ reflects the presence of few edges among the
observable nodes of $\mathcal G$,
and
${\rm{supp}}({\bf\Phi}_S)$ can retrieve
conditional dependence among the observable variables \cite{Zorzi2016_ARlatent}.
The conditional dependence relations among the observable
variables are mainly through the latent variables.
Since $l\leq n$ and ${\rm{rank}}({\bf\Phi}_L)=l$,
the rank of the low-rank part ${\bf\Phi}_L$ can
measure the number of latent variables chosen to model the statistical conditional dependencies of the data \cite{Zorzi2016_ARlatent}.
\end{Remark}
Similar identification problems of (\ref{Phi_SPhi_L}) have been addressed in \cite{Zorzi2016_ARlatent,Ciccone2020_ARlatent,Veedu2022_LDM} under some sparse plus low-rank decomposition
optimization frameworks.
In our problem setting,
since
the low-rank part ${\bf\Phi}_L$ has the dynamical  matrix polynomial form
(\ref{Phi_L})--(\ref{Lambda}),
the matricial reparametrization of the decomposition
optimization problem proposed in \cite{Zorzi2016_ARlatent,Ciccone2020_ARlatent}
cannot be directly implemented.
The work in \cite{Veedu2022_LDM} considers the exact reconstruction problem of ${\rm{supp}}({\bf\Phi}_S)$ and ${\rm{supp}}({\bf\Phi}_L)$ from true inverse spectrum without
an emphasis on the finiteness of the data available.
In this paper,
we aim to propose a new sparse plus low-rank
optimization framework to identify the graphical structure $\mathcal G$ as well as model parameters of (\ref{AR_dyl}) based on (\ref{AR_dyl_PDM}) and (\ref{Phi_SPhi_L})
according to finite observation data ${\boldsymbol{y}}(t)$.

\section{Main Results}
\label{sec:Results}
Define the autocovariance sequence of the \textcolor{blue}{AR Gaussian processes} with latent variables (\ref{AR_dyl}) as
\begin{eqnarray}
{\boldsymbol{R}}_k=\mathbb{E}{\boldsymbol{y}}(t+k){\boldsymbol{y}}(t)^{\T},\non
\end{eqnarray}
Note that ${\boldsymbol{R}}_{-k}={\boldsymbol{R}}_k^{\T}$ since ${\boldsymbol{y}}(t)$ is real.
The corresponding power spectral density matrix is
\begin{eqnarray}
{\bf{\Phi}}_y(e^{{\rm{j}} \omega})=\sum_{k=-\infty}^{\infty}{\boldsymbol{R}}_k e^{-{\rm{j}} k\omega}.\non
\end{eqnarray}
If we have a sequence of observations from $t=1$ to $t=N$,
the sample estimates of the autocovariance matrices can be empirically computed by
\begin{eqnarray}
\label{bfR_k}
\hat{{\boldsymbol{R}}}_k=\frac{1}{N}\sum_{t=1}^{N-k}{\boldsymbol{y}}(t+k){\boldsymbol{y}}(t)^{\T},\ \ k=0,1,\cdots,p_1.
\end{eqnarray}
The identification method is divided into two parts.
One part constructs a sparse plus low-rank
optimization model to estimate the graph topology $\mathcal E$ and AR parameters ${\boldsymbol{A}}_j\ (j=1,2,\cdots,p_1)$.
The other part identifies parameter matrices ${\boldsymbol{W}}_{L,i}\ (i=0,1,\cdots,p_2)$ of the latent variables ${\boldsymbol{x}}(t)$ by solving a constrained rank minimization problem based on AR estimates $\hat{{\boldsymbol{A}}}_j$.
Finally, the identified models are discriminated and selected by some score function.

\subsection{Joint Estimation of Graph Topology and AR Parameters}
\label{sec:ResultsAR}
From (\ref{AR_dyl_PDM}) we have
\begin{eqnarray}
{\bf\Phi}_{W_L}(z)={\bfA}(z){\bf\Phi}_y(z){\bfA}^\ast(z)-{\boldsymbol{I}}_n.\non
\end{eqnarray}
Since ${\bf\Phi}_{W_L}$ is a low-rank matrix with rank $l$,
the corresponding term ${\bfA}(z){\bf\Phi}_y(z){\bfA}^\ast(z)-{\boldsymbol{I}}_n$ is also low-rank.
In addition,
according to (\ref{Phi_S}) and analysis in Remark \ref{Remark2},
the sparsity pattern of the matrix pseudo-polynomial
${\bfA}^\ast(z){\bfA}(z)\in{\mathcal Y}(n,p_1)$
characterizes the graphical structure among the
observable nodes.
Therefore,
an optimization problem with regard to AR parameter matrix polynomial ${\bfA}(z)$ can be constructed as
\textcolor{blue}{
\begin{eqnarray}
\label{S_L_primal}
\mathop{\rm{min}}\limits_{{\bfA}(z)}{\rm{rank}} (\bfA(z){\bf\Phi}_{y}(z)\bfA^{*}(z)-\bfI_n)
+\gamma\|\bfA^{*}(z)\bfA(z)\|_0,\non\\
\end{eqnarray}}
where $\gamma>0$ is a fixed penalty, selected a priori.
Different from directly using the sparse plus low-rank structure of the inverse spectrum associated to observable variables as in \cite{Zorzi2016_ARlatent,Ciccone2020_ARlatent,Veedu2022_LDM},
the
optimization program (\ref{S_L_primal}) is constructed based on the sparse and low-rank structure related to the graphical AR part,
from which the AR parameters and conditional independence relations among the observable \textcolor{blue}{Gaussian variables}
can be obtained simultaneously.
\textcolor{blue}{By introducing two convex penalty functions ${\phi}_{\ast}$ and ${\phi}_{1}$ as surrogates for rank and $l_0$ norm, respectively,
we can obtain a tractable convex relaxation for (\ref{S_L_primal}):}
\begin{eqnarray}
\label{S_L0}
\mathop{\rm{min}}\limits_{{\bfA}(z)}{\phi}_{\ast} [\bfA(z){\bf\Phi}_{y}(z)\bfA^{*}(z)-\bfI_n]
+\gamma{\phi}_{1}[\bfA^{*}(z)\bfA(z)].
\end{eqnarray}
As introduced in \cite{Zorzi2016_ARlatent},
the low-rank penalty function ${\phi}_{\ast}$ can be chosen as
\begin{eqnarray}
\label{phi_ast}
\begin{split}
&{\phi}_{\ast}[\bfA(z){\bf\Phi}_{y}(z)\bfA^{*}(z)-\bfI_n]\\
=&
{\rm{tr}}\left(\frac{1}{2\pi}\int_{-\pi}^{\pi}\bfA(e^{{\rm{j}} \omega}){\bf\Phi}_{y}(e^{{\rm{j}} \omega})\bfA^{*}(e^{{\rm{j}} \omega})-\bfI_nd\omega\right)\\
=&{\rm{tr}}({\bftheta_A{\boldsymbol{K}}{\bftheta}_A^{\T}}-\bfI_n)\\
=&{\rm{tr}}({\boldsymbol{K}}\bfX)-n,
\end{split}
\end{eqnarray}
where
\begin{eqnarray}
\label{boldsymbolK}
           {\boldsymbol{K}}&:=&\setlength\arraycolsep{1pt}
        \left[\begin{array}{cccc}
           {\boldsymbol{R}}_0     &  {\boldsymbol{R}}_1        & \cdots & {\boldsymbol{R}}_{p_1} \\
             {\boldsymbol{R}}_1^{\T}        &  {\boldsymbol{R}}_0       &\cdots  & {\boldsymbol{R}}_{p_1-1}\\
             \vdots    &\vdots    & \ddots & \vdots\\
           {\boldsymbol{R}}_{p_1}^{\T}         & {\boldsymbol{R}}_{p_1-1}^{\T}    & \cdots      & {\boldsymbol{R}}_0 \end{array}\right]\in\mathbb{S}^{n(p_1+1)},\\
  \label{theta_A}
           \bftheta_A&:=&[\bfI_n,{\boldsymbol{A}}_1,{\boldsymbol{A}}_2,\cdots,{\boldsymbol{A}}_{p_1}]\in\mathbb{R}^{n\times n(p_1+1)},\\
  \label{bfX}
           \bfX&=&\bftheta_A^{\T}\bftheta_A.
\end{eqnarray}
The variable $\bfX\in\mathbb{S}^{n(p_1+1)}$ is introduced as a new optimization variable for (\ref{S_L0}),
and is also used for the definition of the sparse penalty function ${\phi}_{1}$.
In view of (\ref{theta_A}) and (\ref{bfX}),
the parameter matrix $\bfX$ \textcolor{blue}{is positive semi-definite and can be partitioned in the form of (\ref{bfX_partition})
with $\bfS$ and $p$ replaced by $\bfX$ and $p_1$, respectively.
We can also obtain that $\bfX_{0,0}=\bfI_n$.}
Based on (\ref{Phi_S}),
the sparse matrix ${\bf\Phi}_S(z)$ can be further written as
\textcolor{blue}{
\begin{eqnarray}
\label{S_L1}
{\bf\Phi}_S(z)={\bfA}^\ast(z){\bfA}(z)={\mathcal R}\bigg\{\sum_{j=0}^{p_1}z^{-j}\bfQ_j \bigg\}.\non
\end{eqnarray}
From (\ref{bfX}), coefficients $\bfQ_j$ can be represented by $\bfX$ as
\begin{eqnarray}
\label{Q_k}
\bfQ_0=\sum_{v=0}^{p_1}\bfX_{v,v},\ \bfQ_j=2\sum_{v=0}^{p_1-j}\bfX_{v,v+j},\ j=1,\cdots,p_1.\non
\end{eqnarray}
Let  the set of matrices be $\bfQ:=[\bfQ_0\ \bfQ_1\ \cdots\ \bfQ_p]\in \mathbb{M}^{n,p_1}$.
It follows that $\bfQ=\mathcal{D}(\bfX)$.
As shown in \cite{Songsiri2010_TSAR,Songsiri2010_gmar},
since $[{\bf\Phi}_S(z)]_{kq}=0$ if and only if the $kq$-th and $qk$-th entries of $\bfQ_j$ are zero for $j=0,\cdots,p_1$,
the sparsity penalty function ${\phi}_{1}$ is defined as
 \begin{eqnarray}
 \begin{split}
 \label{phi1}
 {\phi}_{1}[\bfA^{*}(z)\bfA(z)]&=h_\infty(\bfQ)=h_\infty(\mathcal{D}(\bfX))\\
&={\sum_{q>k}}\mathop{\rm{max}}\limits_{j=0,\cdots,p_1}\bigg\{\vert[\bfQ_j]_{kq}\vert,\vert[\bfQ_j]_{qk}\vert\bigg\},\\
& k=1,\cdots,n-1,\ q=k+1,\cdots,n,
\end{split}
\end{eqnarray}
that is, a sum of the $l_\infty$-norms of vectors of $kq$-th and $qk$-th entries in $\bfQ_j$.
Substituting (\ref{phi_ast}) and (\ref{phi1}) into (\ref{S_L0}), 
we can obtain the matrix parametrization of the optimization problem (\ref{S_L0}):
\begin{eqnarray}
\label{S_L1}
\begin{split}
&\mathop{\rm{min}}\limits_{{\bfX}}\ {\rm{tr}}({\boldsymbol{K}}\bfX)-n+
\gamma h_\infty(\mathcal{D}(\bfX)),\\
&{\rm{s.t.}}\ \ \ \bfX \geq 0,\ \ \bfX_{0,0}=\bfI_n,
\end{split}
\end{eqnarray}
\textcolor{blue}{which looks like the least-squares paradigm augmented
with a sparsity-inducing term  and is similar to
the regularized maximum likelihood estimation problem \cite{Songsiri2010_TSAR,Songsiri2010_gmar}.} 
\begin{Proposition}
If $\bfK>0$,
the optimization problem (\ref{S_L1}) admits a unique solution $\bfX^\circ$ which
can be factorized as $\bfX^\circ=\hat{\bftheta}_A^{\T}\hat{\bftheta}_A$ with $\hat{\bftheta}_A:=[{\bfI}_n,\hat{\bfA}_1,\cdots,\hat{\bfA}_{p_1}]\in\mathbb{R}^{n\times n(p_1+1)}$.
\end{Proposition}
The proof is provided in Appendix A by exploiting duality
theory.}

Taking $\lambda:=\frac{\gamma}{1+\gamma}$,
the convex program (\ref{S_L1}) is equivalent to the following
formulation:
\textcolor{blue}{
\begin{eqnarray}
\label{S_L2}
 \begin{split}
&\mathop{\rm{min}}\limits_{{\bfX}}\ (1-\lambda)[{\rm{tr}}({\boldsymbol{K}}\bfX)-n]+
\lambda h_\infty(\mathcal{D}(\bfX)),\\
&{\rm{s.t.}}\ \ \ \bfX \geq 0,\ \ \bfX_{0,0}=\bfI_n.
\end{split}
\end{eqnarray}}
where the penalty parameter $\lambda$ will be chosen from the range $(0,1)$ via the cross-validation method \cite{Friedman2008_Glasso},
and the best value of $\lambda$ is selected according to some model fitness function.
\textcolor{blue}{
Common convex optimization methods such as interior-point methods and gradient projection algorithms can be applied to solve (\ref{S_L2}) \cite{Boyd2004_CO}.
In practice,
the matrix $\boldsymbol{K}$ in (\ref{S_L2}) is replaced by $\hat{\boldsymbol{K}}$ consisting of covariance matrix estimates $\hat{{\boldsymbol{R}}}_k$ computed by (\ref{bfR_k}).
In this case, the estimate $\hat{\boldsymbol{K}}$ is block-Toeplitz, positive definite, and satisfies the hypothesis in Proposition 4.}
As a consequence,
given a string of penalty coefficients $\lambda$,
sparse solutions $\hat{\bfX}$
can be obtained by solving
 the optimization problem (\ref{S_L2}),
 following the AR parameter estimates $\hat\bfA_1,\cdots,\hat\bfA_{p_1}$ determined from $\hat{\bfX}$ by spectral decomposition and the sparse part of inverse spectrums computed by
 \begin{eqnarray}
 \label{hatbfPhi_S}
\hat{{\bf\Phi}}_S(z)&=&{\mathcal R}\bigg\{\sum_{j=0}^{p_1}z^{-j}\hat{\bfQ}_j \bigg\},\\
\hat{\bfQ}_0&=&\sum_{v=0}^{p_1}\hat{\bfX}_{v,v},\ \hat{\bfQ}_j=2\sum_{v=0}^{p_1-j}\hat{\bfX}_{v,v+j},\ j=1,\cdots,p_1.\non
\end{eqnarray}
We further normalize $\hat{{\bf\Phi}}_S$ and filter out the elements smaller than the threshold set a priori (see more details in \cite{You2022_GME_ARMA}).
Then the sparsity pattern of new $\hat{{\bf\Phi}}_S$ reconstructs the graph topology among observable nodes and reveals the positions of non-zero entries in the parameter matrix ${\bftheta}_A$.

\textcolor{blue}{In the optimization model (\ref{S_L2}),
the trace norm of ${\boldsymbol{K}}\bfX$ is used as a convex heuristic for low-rank recovery.
For the low-rank penalty function ${\phi}_{\ast}$ in (\ref{phi_ast}),
although a low-rank term ${\boldsymbol{K}}\hat{\bfX}$ can be obtained with the solution $\hat{\bfX}=\hat{\bftheta}_A^{\T}\hat{\bftheta}_A$,
the low-rank property of the term $\hat{\bftheta}_A{\boldsymbol{K}}\hat{\bftheta}_A^{\T}-\bfI_n$ cannot be guaranteed.
As a result, the estimates of AR parameters may not be reliable.
Nevertheless, the graph topology ${\rm{supp}}(\boldsymbol{\Phi}_S)$ can be estimated with an appropriate penalty parameter $\lambda$.
This is due to the fact that ${\bf\Phi}_S$ is determined by the graphical AR part instead of hidden variables.
In the case that there
exists small estimation error in graph topology,
although with some element wise mismatch of non-zero entries in AR parameter estimates,
the structural pattern of the sparse matrix ${\bf\Phi}_S$ can be  almost preserved.} 
To further improve the AR-parameter estimation accuracy,
we take
the graph topology estimate ${\rm{supp}}(\hat{{\bf\Phi}}_S)$ obtained from (\ref{S_L2}) as
the sparsity constraint to construct the matrix rank minimization problem:
\begin{eqnarray}
 \begin{split}
\label{rankm1}
& \mathop{\rm{min}}\limits_{{\bftheta}_A}\ {\rm{rank}}({\bftheta_A{{\boldsymbol{K}}}{\bftheta}_A^{\T}}-\bfI_n),\\
&{\rm{s.t.}} \ \ \ \bftheta_A(:,jn+1:jn+n)\in {\rm{supp}}(\hat{{\bf\Phi}}_S),\ j=1,\cdots,p_1,\\
&\ \ \ \ \ \ \ \ \bftheta_A(:,1:n)=\bfI_n.
\end{split}
\end{eqnarray}
Since the problem (\ref{rankm1}) is a constrained quadratic optimization problem with the variable ${\bftheta}_A$ coupled together,
it cannot be directly solved.
To deal with this difficulty,
an equivalent reformulation of (\ref{rankm1}) is introduced by using the Schur complement lemma \cite{Ouellette1981_SC}:
\begin{eqnarray}
 \begin{split}
\label{rankm2}
& \mathop{\rm{min}}\limits_{{\bftheta}_A}\ {\rm{rank}}
\setlength\arraycolsep{2pt}
        \left[\begin{array}{cc}
          {\boldsymbol{K}}^{-1}  &  {\bftheta}_A^{\T}   \\
             \bftheta_A       &  \bfI_n     \end{array}\right],\\
&{\rm{s.t.}} \ \ \ \bftheta_A(:,jn+1:jn+n)\in {\rm{supp}}(\hat{{\bf\Phi}}_S),\ j=1,\cdots,p_1,\\
&\ \ \ \ \ \ \ \ \bftheta_A(:,1:n)=\bfI_n.\non
\end{split}
\end{eqnarray}
\textcolor{blue}{
Define the transformed objective matrix as
\begin{eqnarray}
\bfX_L(\bftheta_A):=\left[\begin{array}{cc}
           {\boldsymbol{K}}^{-1}  &  {\bftheta}_A^{\T}   \\
             \bftheta_A       &  \bfI_n     \end{array}\right]\in\mathbb{S}^{n(p_1+2)}.\non
\end{eqnarray}
The nuclear-norm heuristic of the above rank minimization problem can be written as
\begin{eqnarray}
 \begin{split}
\label{rankm2}
& \mathop{\rm{min}}\limits_{{\bftheta}_A}\ \|
\bfX_L(\bftheta_A)\|_\ast,\\
&{\rm{s.t.}} \ \ \ \bftheta_A(:,jn+1:jn+n)\in {\rm{supp}}(\hat{{\bf\Phi}}_S),\ j=1,\cdots,p_1,\\
&\ \ \ \ \ \ \ \ \bftheta_A(:,1:n)=\bfI_n.
\end{split}
\end{eqnarray}
The singular value decomposition (SVD) of $\bfX_L(\bftheta_A)$ is
\begin{eqnarray}
\bfX_L(\bftheta_A)&=&\bfU_{\bfX_L}\Sigma_{\bfX_L}\bfV_{\bfX_L}^{\T}=\sum_{k=1}^{r}\sigma_{\bfX_L,k}\bfu_{\bfX_L,k}\bfv_{\bfX_L,k}^{\T},\non
\end{eqnarray}
where $r={\rm{rank}}(\bfX_L(\bftheta_A))$. 
By the Schur complement lemma, we have
\begin{eqnarray}
r={\rm{rank}}({\bftheta_A{{\boldsymbol{K}}}{\bftheta}_A^{\T}}-\bfI_n)+{\rm{rank}}({{\boldsymbol{K}}})
=l(p_2+1)+n(p_1+1)\non
\end{eqnarray}
if $l(p_2+1)<n$.
Denote the column and row spaces of $\bfX_L(\bftheta_A)$ by $U_{\bfX_L}$ and $V_{\bfX_L}$, respectively, and denote $\tilde{n}=n(p_1+2)$.
\begin{Definition}
Let $U$ be a subspace of $\mathbb{R}^{\tilde{n}}$ of dimension $r$
and ${{\mathbf{P}}}_u$ be the orthogonal projection onto $U$.
Then the coherence of $U$ (vis-$\grave{a}$-vis the standard basis (${\mathbf{e}}_i$)) is defined to be
$\mu(U)\equiv \frac{\tilde{n}}{r}\mathop{\rm{max}}\limits_{1\leq i\leq \tilde{n}}\|{{\mathbf{P}}}_U{\mathbf{e}}_i\|^2$.
\end{Definition}
\begin{Theorem}
\label{Theorem1}
Assume that $\bfX_L(\bftheta_A)$ obey two assumptions:\\ 
${\bfA0}$ The coherences obey ${\rm{max}}(\mu(U_{\bfX_L}),\mu(V_{\bfX_L}))\leq \mu_0$ for some
$\mu_0>0$;\\
${\bfA1}$ The $\tilde{n}\times \tilde{n}$ matrix $\sum_{1\leq k\leq r}\bfu_{\bfX_L,k}\bfv_{\bfX_L,k}^{\T}$ has a maximum entry bounded by $\mu_1\sqrt{r/\tilde{n}^2}$  in absolute value for some $\mu_1>0$.\\
Suppose we have obtained effective graph topology estimate ${\rm{supp}}(\hat{{\bf\Phi}}_S)$, that is ${\rm{supp}}(\hat{{\bf\Phi}}_S)$ is close to or equal to the sparsity pattern of true ${\bf\Phi}_S$ in (\ref{Phi_S}).
In this case totally
 $m=\tilde{n}^2-2p_1\vert{\rm{supp}}(\hat{{\bf\Phi}}_S)\vert$ entries of $\bfX_L(\bftheta_A)$ are known.
 Then there exist constants $C$, $c$ such that if
\begin{eqnarray}
m\geq C{\rm{max}}(\mu_1^2,\mu_0^{1/2}\mu_1,\mu_0\tilde{n}^{1/4})\tilde{n}r(\beta \log{\tilde{n}})\non
\end{eqnarray}
for some $\beta>2$, then the optimal value $\hat{\bftheta}_A$ of the problem (\ref{rankm2})
 is unique
and equal to
the true underlying matrix ${\bftheta}_A$ in (\ref{theta_A}) with probability at least $1-c\tilde{n}^{-\beta}$. For $r\leq \mu_0^{-1}\tilde{n}^{1/5}$
this estimate can be improved to
\begin{eqnarray}
m\geq C\mu_0\tilde{n}^{6/5}r(\beta \log{\tilde{n}})\non
\end{eqnarray}
with the same probability of success.
\end{Theorem}
The demonstration is provided in Appendix B based on the theoretical results in \cite{Cand2009_MC}.}

\textcolor{blue}{Theorem \ref{Theorem1} implies that if the entries of $\bfX_L(\bftheta_A)$ available including the subblock matrix $\bfK$ and zero entries of $\bftheta_A$ characterized by the identified graph topology ${\rm{supp}}(\hat{{\bf\Phi}}_S)$ are relatively accurate,
effective AR parameter estimates can be obtained from (\ref{rankm2}) with high probability.
 Note that the block-Toeplitz matrix estimate $\hat{\boldsymbol{K}}$ is used to replace $\boldsymbol{K}$ in computation,
which may result in estimation error.
Inspired by \cite{Fazel2003_Logdet,Mohane2010_RN},
the reweighted trace heuristic can be applied as improving on the solution of the nuclear norm heuristic (\ref{rankm2}),
which solves a semidefinite program at each iteration and can find a local minimum.
The computation steps are as follows:}
\begin{eqnarray}
\label{bfthetaAk1}
\hat{\bftheta}_A^{k+1}&=&{\rm{arg min}}\|{\bfW}_1^k\bfX_L(\bftheta_A){\bfW}_2^k\|_\ast,\\
&&{\rm{s.t.}} \bftheta_A(:,jn+1:jn+n)\in {\rm{supp}}(\hat{{\bf\Phi}}_S),\ j=1,\cdots,p_1,\non\\
&&\ \ \ \ \bftheta_A(:,1:n)=\bfI_n,\non\\
\label{bfTk1}
\bfT^{k+1}&=&({\bfW}_1^k)^{-1}\bfU\Sigma\bfU^{\T}({\bfW}_1^k)^{-1},\\
\label{bfZk1}
\bfZ^{k+1}&=&({\bfW}_2^k)^{-1}\bfV\Sigma\bfV^{\T}({\bfW}_2^k)^{-1},\\
\label{bfW1k1}
{\bfW}_1^{k+1}&=&(\bfT^{k+1}+\epsilon\bfI_{n(p_1+2)})^{-1/2},\\
\label{bfW2k1}
{\bfW}_2^{k+1}&=&(\bfZ^{k+1}+\epsilon\bfI_{n(p_1+2)})^{-1/2},
\end{eqnarray}
where the nuclear norm is characterized by
\begin{eqnarray}
\|\bfX_L\|_\ast&=&\frac{1}{2}\mathop{\rm{min}}\limits_{\bfX_L}\
({\rm{tr}}\bfT+{\rm{tr}}\bfZ),\non\\
&&{\rm{s.t.}}\left[\begin{array}{cc}
           \bfT  &  \bfX_L   \\
             \bfX_L^{\T}       &  \bfZ     \end{array}\right]\geq 0,\non
\end{eqnarray}
the symbol $\epsilon>0$ is a small regularization constant,
and ${\bfW}_1^k\bfX_L^{k+1}{\bfW}_2^k=\bfU\Sigma\bfV^{\T}$
is the reduced SVD of
${\bfW}_1^k\bfX_L^{k+1}{\bfW}_2^k$. 
\textcolor{blue}{The initial values of ${\bfW}_1^{0}$ and ${\bfW}_2^{0}$ are often taken as identity matrices,
so that the first iteration only needs to minimize ${\rm{tr}}\bfT+{\rm{tr}}\bfZ$ subject to graph topology conditions,
which is equivalent to (\ref{rankm2}).
The iterations that follow can reduce the rank of $\bfX_L(\bftheta_A)$ further and converge to a stationary solution $\hat{\bftheta}_A$.}

\subsection{Identification of latent-variable parameters}
\label{sec:Resultslatent}
According to the AR parameter estimate
$\hat{\bftheta}_A:=[\bfI_n,\hat{{\boldsymbol{A}}}_1,\cdots,\hat{{\boldsymbol{A}}}_{p_1}]$
obtained from (\ref{rankm2}),
the corresponding parameter matrix polynomial is
\begin{eqnarray}
\label{hatbfAz}
\hat{{\bfA}}(z):={\boldsymbol{I}_n}+\hat{{\bfA}}_1z^{-1}+\hat{{\bfA}}_2z^{-2}+\cdots+\hat{{\bfA}}_{p_1}z^{-p_1}.
\end{eqnarray}
Let $\bfy_{AR}(t)$ be the finite length trajectory obtained by passing through the filter $\hat{{\bfA}}(z)$ the
trajectory $\bfy(t)$ with zero initial conditions:
\begin{eqnarray}
\label{yAR}
\bfy_{AR}(t):=\hat{{\bfA}}(z)\bfy(t).
\end{eqnarray}
Then the sample covariance matrices of the \textcolor{blue}{Gaussian process} (\ref{yAR}) can be computed by
\begin{eqnarray}
\label{hatRkAR}
\hat{{\boldsymbol{R}}}_k^{AR}=\frac{1}{N}\sum_{t=1}^{N-k}{\boldsymbol{y}_{AR}}(t+k){\boldsymbol{y}_{AR}}(t)^{\T},\ \ k=0,1,\cdots,p_2,
\end{eqnarray}
and the estimate $\hat{{\bf\Phi}}_{AR}$ of the AR spectral density matrix  ${\bf\Phi}_{AR}(z):={\bfA}(z){\bf\Phi}_y(z){\bfA}^\ast(z)$
can be obtained by the truncated
periodogram:
\begin{eqnarray}
\label{hatbfPhiAR}
\hat{{\bf\Phi}}_{AR}(e^{{\rm{j}} \omega})=\sum_{k=-p_2}^{p_2}\hat{{\boldsymbol{R}}}_k^{AR} e^{-{\rm{j}} k\omega},\ \hat{{\boldsymbol{R}}}_{-k}^{AR} =[\hat{{\boldsymbol{R}}}_k^{AR}]^{\T}.
\end{eqnarray}
 Some windowing methods can be used to smooth $\hat{{\bf\Phi}}_{AR}$ to ensure that $\hat{{\bf\Phi}}_{AR}(e^{{\rm{j}}\omega})\geq 0$ for all
 $\omega \in [-\pi,\pi]$ \cite{Stoica2005_Spectral}.
Based on (\ref{AR_dyl_PDM}), it can be derived that
\begin{eqnarray}
\label{AR_dyl_PDM3}
{\bf\Phi}_{AR}
=&{\bf\Phi}_{W_L}+{\boldsymbol{I}}_n,\non
\end{eqnarray}
which can be seen as the simplified 'low-rank plus diagonal' structure decomposition of the spectrum ${\bf\Phi}_{AR}$ studied in \cite{Falconi2021_MA_factor,Falconi2021_ARMA_factor}
with the diagonal part known a priori as an identity matrix.
Therefore,
 we \textcolor{blue}{simplify} the optimization framework constructed in \cite{Falconi2021_MA_factor,Falconi2021_ARMA_factor},
and estimate ${\bf\Phi}_{W_L}$ by solving the following optimization problem:
\begin{eqnarray}
\label{PhiWL_opt}
\begin{split}
&\mathop{\rm{min}}\limits_{{\bf\Phi}_{W_L}}\
{\rm{tr}}\left(\frac{1}{2\pi}\int_{-\pi}^{\pi}{\bf\Phi}_{W_L}(e^{{\rm{j}} \omega})d\omega\right),\\
&\ {\rm{s.t.}} \ \ \ \|{\bf\Phi}_{W_L}+{\boldsymbol{I}}_n-\hat{{\bf\Phi}}_{AR}\|_F\leq \delta,\ {\bf\Phi}_{W_L}\geq 0,
\end{split}
\end{eqnarray}
where $\delta>0$ represents
a proxy for the uncertainty on the estimate $\hat{{\bf\Phi}}_{AR}$.
Analogous to the divergence constraint introduced in
\cite{Ciccone2020_ARlatent,Falconi2021_MA_factor,Falconi2021_ARMA_factor,Ciccone2019_FMR,Crescente2020_AR_factor},
the Frobenius norm constraint in (\ref{PhiWL_opt})
imposes that true spectral density ${\bf\Phi}_{AR}$ belongs to a set 'centered' in the nominal spectrum estimate $\hat{{\bf\Phi}}_{AR}$ with prescribed tolerance $\delta$.
With this convex constraint,
the simplified optimization problem (\ref{PhiWL_opt}) is a tractable convex program.

Next, we reparametrize (\ref{PhiWL_opt}) in a matricial form
for the sake of intuitive parameter estimation.
Define the latent parameter matrix to be identified as
\begin{eqnarray}
\label{theta_l}
           \bftheta_l:=[{\boldsymbol{W}}_{L,0}^{\T},{\boldsymbol{W}}_{L,1}^{\T},\cdots,{\boldsymbol{W}}_{L,p_2}^{\T}]^{\T}\in\mathbb{R}^{n(p_2+1)\times l}.\non
\end{eqnarray}
Then ${\bf\Phi}_{W_L}$ can be expressed by
\begin{eqnarray}
\label{bfPhiWLDelta}
{\bf\Phi}_{W_L}(e^{{\rm{j}} \omega})&=&\Delta(e^{{\rm{j}} \omega})\bfL\Delta^\ast(e^{{\rm{j}} \omega}),\\
\Delta (e^{{\rm{j}} \omega})&:=&[\bfI_n,e^{-{\rm{j}} \omega}\bfI_n,\cdots,e^{-{\rm{j}}p_2  \omega}\bfI_n]\in\mathbb{R}^{n\times n(p_2+1)},\non\\
\label{bfLtheta}
\bfL&=&\bftheta_l\bftheta_l^{\T}.
\end{eqnarray}
Note that the matrix $\bfL$ inherits the low-rank property of ${\bf\Phi}_{W_L}$ with ${\rm{rank}}(\bfL)=l$,
and the condition ${\bf\Phi}_{W_L}\geq 0$ can be replaced with $\bfL\geq 0$ \cite{Zorzi2016_ARlatent}.
The optimization objective function in (\ref{PhiWL_opt})
is thereby written as
\begin{eqnarray}
\label{bfL_opm}
&&{\rm{tr}}\left(\frac{1}{2\pi}\int_{-\pi}^{\pi}{\bf\Phi}_{W_L}(e^{{\rm{j}} \omega})d\omega\right)\non\\
=&&
{\rm{tr}}\left(\frac{1}{2\pi}\int_{-\pi}^{\pi}\Delta(e^{{\rm{j}} \omega})\bfL\Delta^\ast(e^{{\rm{j}} \omega}) d\omega\right)\non\\
=&&
{\rm{tr}}\left(\bfL\frac{1}{2\pi}\int_{-\pi}^{\pi}\Delta^\ast(e^{{\rm{j}} \omega})\Delta(e^{{\rm{j}} \omega}) d\omega\right)\non\\
=&&{\rm{tr}}\bfL.
\end{eqnarray}
By comparing (\ref{hatbfPhiAR}) with (\ref{bfPhiWLDelta}),
the Frobenius norm constraint in (\ref{PhiWL_opt}) can
be split into
\begin{eqnarray}
\label{PhiWL_split}
\left\{
\begin{aligned} &\|\sum_{v=0}^{p_2}\bfL_{v,v}+{\boldsymbol{I}}_n-\hat{{\boldsymbol{R}}}_0^{AR}\|_F\leq \delta_0,\\
&\|\sum_{v=0}^{p_2-1}\bfL_{v,v+1}^{\T}-\hat{{\boldsymbol{R}}}_1^{AR}\|_F\leq \delta_1,\\
&\|\sum_{v=0}^{p_2-2}\bfL_{v,v+2}^{\T}-\hat{{\boldsymbol{R}}}_2^{AR}\|_F\leq \delta_2,\\
&\ \ \ \ \ \ \ \ \ \ \ \ \ \ \ \vdots\\
&\|\bfL_{0,p_2}^{\T}-\hat{{\boldsymbol{R}}}_{p_2}^{AR}\|_F\leq \delta_{p_2},
\end{aligned}
\right.
\end{eqnarray}
where
 $\bfL_{i,i}\in\mathbb{S}^{n}$ and
$\bfL_{i,j}\in\mathbb{R}^{n\times n}$ $(i\neq j)$ denote
 the subblocks of $\bfL\in\mathbb{S}^{n(p_2+1)}$ partitioned in the form of (\ref{bfX_partition})
with $\bfS$ and $p$ replaced by $\bfL$ and $p_2$, respectively.
The tolerance parameters $\delta_0,\cdots,\delta_{p_2}$
reflecting the accuracy of the estimates $\hat{{\boldsymbol{R}}}_k^{AR}\ (\ k=0,1,\cdots,p_2)$
can be chosen by a resampling-based method \cite{Falconi2021_ARMA_factor}.
Define the resampled process $\hat{\bfy}_{AR}=\{\hat{\bfy}_{AR}(t):t \in {\mathbb Z}\}$
as
 \begin{eqnarray}
 \hat{\bfy}_{AR}(t)&:=&\bfW(z)\bfe(t),\non\\
 \bfW(z)&:=&{\boldsymbol{W}}_{0}+{\boldsymbol{W}}_{1}z^{-1}+\cdots+{\boldsymbol{W}}_{p_2}z^{-p_2},\non
 \end{eqnarray}
 where $\bfW(z)$
 is the minimum phase
spectral factor of $\hat{{\bf\Phi}}_{AR}$ satisfying
$\hat{{\bf\Phi}}_{AR}(z)=\bfW(z)\bfW^{\ast}(z)$,
and $\bfe(t)$ is the normalized white Gaussian noise process.
Given a realization of $\hat{\bfY}_{AR}=\{\hat{\bfy}_{AR}(1),\cdots,\hat{\bfy}_{AR}(N)\}$
generated from $\hat{{\bf\Phi}}_{AR}$ and $\bfe(t)$,
we can obtain a realization of the random variable
$\|\hat{{\boldsymbol{R}}}_k^{AR}-\hat{{\boldsymbol{R}}}_{r,k}^{AR}\|_F$,
where
\begin{eqnarray}
\hat{{\boldsymbol{R}}}_{r,k}^{AR}=\frac{1}{N}\sum_{t=1}^{N-k}\hat{\bfy}_{AR}(t+k)\hat{\bfy}_{AR}(t)^{\T},\ \ k=0,1,\cdots,p_2\non
\end{eqnarray}
are resampling covariance estimates.
Consequently,
by choosing a desired probability $\alpha_k\in (0,1)$,
it is possible to compute numerically $\delta_{k,\alpha_k}$ such that ${\rm{Pr}}(\|\hat{{\boldsymbol{R}}}_k^{AR}-\hat{{\boldsymbol{R}}}_{r,k}^{AR}\|_F\leq \delta_{k,\alpha_k})=\alpha_k$ for $k=0,1,\cdots,p_2$ by a standard
Monte Carlo procedure.

By virtue of relations (\ref{bfL_opm}) and (\ref{PhiWL_split}),
the optimization problem (\ref{PhiWL_opt}) is reformulated as
\begin{eqnarray}
\label{PhiWL_opt_marix}
\begin{split}
&\mathop{\rm{min}}\limits_{\bfL}\
{\rm{tr}}\bfL,\\
&\ {\rm{s.t.}} \ \|\sum_{v=0}^{p_2}\bfL_{v,v}+{\boldsymbol{I}}_n-\hat{{\boldsymbol{R}}}_0^{AR}\|_F\leq \delta_0,\\
&\ \ \ \ \ \ \ \|\sum_{v=0}^{p_2-k}\bfL_{v,v+k}^{\T}-\hat{{\boldsymbol{R}}}_k^{AR}\|_F\leq \delta_k,\ k=1,\cdots,p_2,\\
&\ \ \ \ \ \ \ \bfL\geq 0.
\end{split}
\end{eqnarray}
\textcolor{blue}{The work in \cite{Falconi2021_MA_factor,Falconi2021_ARMA_factor} has
proven the existence of the solution of the complex version of the problem (\ref{PhiWL_opt_marix}) via variational analysis.
The simplified convex program (\ref{PhiWL_opt_marix}) can be directly solved by using convex optimization packages such as CVX and yalmip.
The number of latent variables $\hat{l}$ can be determined from the rank of the optimal solution $\hat{\bfL}$.
A minimum-phase factor $\hat{\bftheta}_l$ can be recovered based on (\ref{bfLtheta}) by a computationally efficient spectral factorization approach \cite{Cao2023_Fa_rd},
which exploits a deterministic relation inside the factor and the coprime factorization with an inner factor in computation.}

\subsection{Summary of the Identification Algorithm}
Given a string of penalty parameters $\lambda$,
a set of corresponding \textcolor{blue}{graphical AR Gaussian models} with hidden variables can be estimated based on (\ref{S_L2}),
(\ref{rankm2}) and (\ref{PhiWL_opt_marix}).
Inspired by \cite{Zorzi2016_ARlatent},
a score function is introduced to discriminate among the estimated models:
\begin{eqnarray}
\label{score}
f(\hat{\bfPhi}_y^{NP},\hat{\bfPhi}_{y}^{P},\hat{\bfPhi}_{S},\hat{l})&=&
\mathbb{D}(\hat{\bfPhi}_{y}^{NP}\|\hat{\bfPhi}_{y}^{P})\times p,
\end{eqnarray}
where $\hat{\bfPhi}_{y}^{NP}$ is the smoothed non-parametric spectral estimate of the latent-variable AR processes
(\ref{AR_dyl}) computed from the sampled covariance with different lags (\ref{bfR_k}),
and $\hat{\bfPhi}_{y}^{P}$ is the parametric spectral estimate obtained by
\begin{eqnarray}
\hat{\bfPhi}_{y}^{P}(z)
=\hat{\bfA}(z)^{-1}[\hat{\bf\Phi}_{W_L}(z)+{\boldsymbol{I}}_n]\hat{\bfA}^\ast(z)^{-1}\non
\end{eqnarray}
with $\hat{\bfA}(z)$ defined in (\ref{hatbfAz})
and $\hat{\bf\Phi}_{W_L}(z):=\Delta(z)\hat{\bfL}\Delta^\ast(z)$.
The cost
\begin{eqnarray}
\mathbb{D}(\hat{\bfPhi}_{y}^{NP}\|\hat{\bfPhi}_{y}^{P})&:=&
\frac{1}{2}\Big\{\frac{1}{2\pi}\int_{-\pi}^{\pi}
\Big[{\rm{\log\det}}\Big(\Big(\hat{\bfPhi}_{y}^{NP}\Big)^{-1}\hat{\bfPhi}_{y}^{P}
\Big)\non\\
&&\ \ +\Big\langle \hat{\bfPhi}_{y}^{NP}, \Big(\hat{\bfPhi}_{y}^{P}\Big)^{-1}\Big\rangle\Big] d\omega-n\Big\}\non
\end{eqnarray}
is the relative entropy rate between $\hat{\bfPhi}_{y}^{NP}$ and $\hat{\bfPhi}_{y}^{P}$, ranking the adherence of
$\hat{\bfPhi}_{y}^{P}$ to the data.
The term
\begin{eqnarray}
p&=&\vert{\rm{supp}}(\hat{\bfPhi}_{S})\vert-n+n\hat{l}\non
\end{eqnarray}
is the total number of edges in the latent-variable graphical
model (\ref{AR_dyl}),
which can characterize the complexity of the model.
Subsequently,
the most appropriate model with good data fitness and low model complexity is selected with the lowest
score $f$.

To sum up,
a sparse plus low-rank identification approach is proposed for the \textcolor{blue}{latent-variable graphical
AR Gaussian model (\ref{AR_dyl})},
and the identification procedure is given in Algorithm 1.
\begin{algorithm}[!htb]
\caption{Sparse plus low-rank identification algorithm for Model (\ref{AR_dyl})} \label{Algorithm}
{\bf Input:} Observed system output $\{\bfy(t)$, $t=1,2,\cdots,N\}$.

{\bf Output:}
Graph topology $\hat{\mathcal E}$,
model parameters $\hat\bfA_j$ ($j=1,2,\cdots,p_1$),
$\hat{\boldsymbol{W}}_{L,i}$ ($i=0,1,\cdots,p_2$),
 and the number of  latent variables $\hat{l}$.
\begin{algorithmic}[1]


\STATE
 Construct $\hat{\bfK}$ based on (\ref{boldsymbolK})
with $\hat\bfR_k$ computed from (\ref{bfR_k}).

\STATE
Given a string of penalty coefficients $\lambda$ chosen
via cross-validation,
determine a list of $\hat\bfX$  with (\ref{S_L2}).

\STATE
Compute $\hat\bfPhi_S$ with (\ref{hatbfPhi_S}),
and obtain $\hat{\mathcal E}$ from the normalized and filtered ${\rm{supp}}(\hat\bfPhi_S)$.

\STATE
 Update $\hat{\bftheta}_A$ iteratively by (\ref{bfthetaAk1})--(\ref{bfW2k1}).

\STATE
Construct $\hat{\bfy}_{AR}(t)$ with (\ref{hatbfAz}) and (\ref{yAR}) and compute $\hat{\bfR}_k^{AR}$ from (\ref{hatRkAR}).

\STATE
Obtain
$\hat{\bfL}$ by solving (\ref{PhiWL_opt_marix}) with $\delta_0,\cdots,\delta_{p_2}$ chosen by the
Monte Carlo technique.

\STATE
Determine $\hat{l}$ from the rank of $\hat{\bfL}$
and recover $\hat{\bftheta}_l$ by spectral factorization.

\STATE
Select the model with  the lowest fitness score (\ref{score}).
\end{algorithmic}
\end{algorithm}


\section{Simulation Example}
\label{sec:Simulation}
In this section, we evaluate the proposed identification approach
on a randomly generated data set and a real data set,
respectively.

\subsection{Simulation Example 1}
Consider a \textcolor{blue}{latent-variable graphical AR Gaussian model}
of orders $p_1=2$, $p_2=1$ with $n=10$ observable variables and $l=1$ latent variables:
\begin{eqnarray}
\label{simu1}
{\boldsymbol{y}}(t)+{\boldsymbol{A}}_1{\boldsymbol{y}}(t-1)+{\boldsymbol{A}}_2{\boldsymbol{y}}(t-2)
&=&{\boldsymbol{W}}_{L,0}{\boldsymbol{x}}(t)\non\\ &&+{\boldsymbol{W}}_{L,1}{\boldsymbol{x}}(t-1)+{\boldsymbol{\omega}}(t),\non
\end{eqnarray}
where $\bfA_j\in\mathbb{R}^{10\times 10}\ (j=1,2)$ are sparse AR parameter matrices with
$[\bfA_1]_{16}=[\bfA_1]_{94}=[\bfA_1]_{10,3}=[\bfA_2]_{25}=[\bfA_2]_{57}=[\bfA_2]_{81}=1$
and $\bfW_{L,i}\in\mathbb{R}^{10\times 1}\ (i=0,1)$ are latent-variable parameter matrices with entries randomly generated within $(0,1)$ such that the whole system is stable.

\begin{figure}[!hbt]
  \centering
  \includegraphics[width=\hsize]{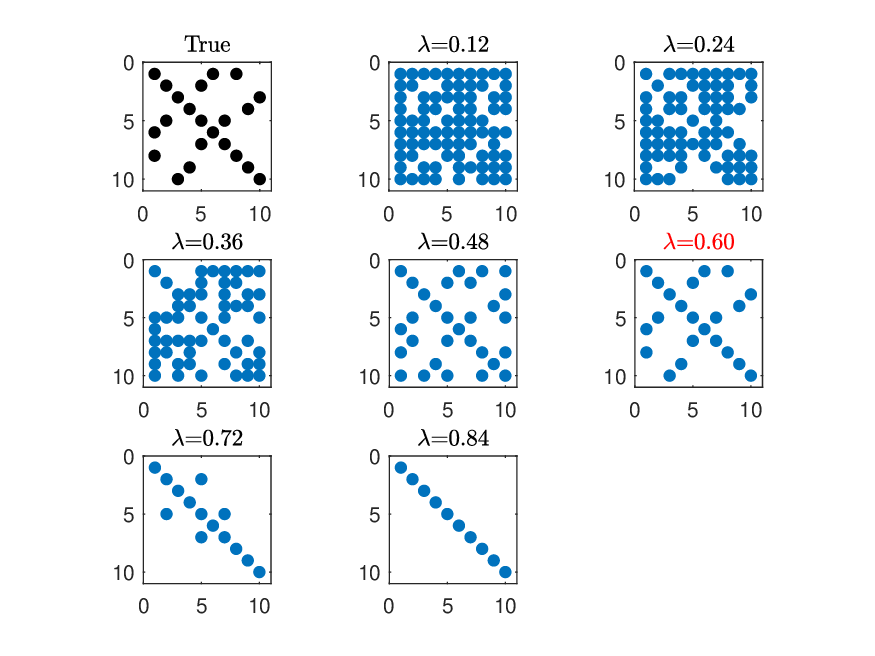}
  \caption{Graph topologies of the true model and optimal models estimated for different penalty parameters $\lambda$}
  \label{Fig_topology}
\end{figure}

In the simulation,
the sample size is taken as $N=5000$,
and the parameters $\delta_{k,\alpha_k}\ (k=0,1)$ are chosen
following the empirical procedure presented in Section~\ref{sec:Resultslatent} for $\alpha_k= 0.95$ from 200
Monte Carlo experiments.
The estimated graph topologies for the observable variables under different penalty parameters $\lambda$ are shown in Fig. \ref{Fig_topology},
where the square represents an edge between two observable nodes.
The number of edges in each model topology,
the corresponding score function $f$ and
estimated number of latent variables $\hat{l}$
are listed in Table~\ref{tab1}.
As the penalty parameter $\lambda$ increases,
the estimated model topology tends to be sparse,
favoring less conditional dependence
relations among the observable variables.
The estimate value $\hat{l}$ does not strictly increase with the increment of $\lambda$ like the identification method  proposed in \cite{Ciccone2020_ARlatent}.
This is because the low-rank matrix $\hat{\bf\Phi}_{W_L}$ whose rank equals to $\hat{l}$ is
 determined by sequentially solving optimization problems
 (\ref{S_L2}),
(\ref{rankm2}) and (\ref{PhiWL_opt}),
where only the optimization  paradigm (\ref{S_L2}) is similar to that in \cite{Ciccone2020_ARlatent}.
With the minimum value of score function $f=18.27$,
the selected model topology for $\lambda=0.60$ is marked in red in Fig. \ref{Fig_topology},
which agrees with the true one.
Also, the number of latent variables is accurately estimated ($\hat{l}=1$) in the selected model.
\begin{table*}[!htb]
  \centering
  \caption{The number of edges in graph topology, the number of latent variables and score function value of optimal models estimated for different penalty parameters $\lambda$}
  \label{tab1}
 \renewcommand{\arraystretch}{1}
\footnotesize \doublerulesep 0.2pt \tabcolsep 13pt 
  \begin{tabular}{lccccccc}\hline
$\lambda$  & 0.12     &  0.24   &  0.36 & 0.48 & 0.60  & 0.72 & 0.84\\\hline
  Number of edges in graph topology &35  &29  &22  &10  &6  & 2    &0\\
  Number of latent variables &2	&1	&1	&1	&1	 &5	&8 \\
  Score function value &65.05	&56.28	&44.73 &24.95	&18.27 &34.55 	&153.22 \\\hline
  \end{tabular}
\end{table*}

\begin{figure}[!hbt]
  \centering
  \includegraphics[width=0.8\hsize]{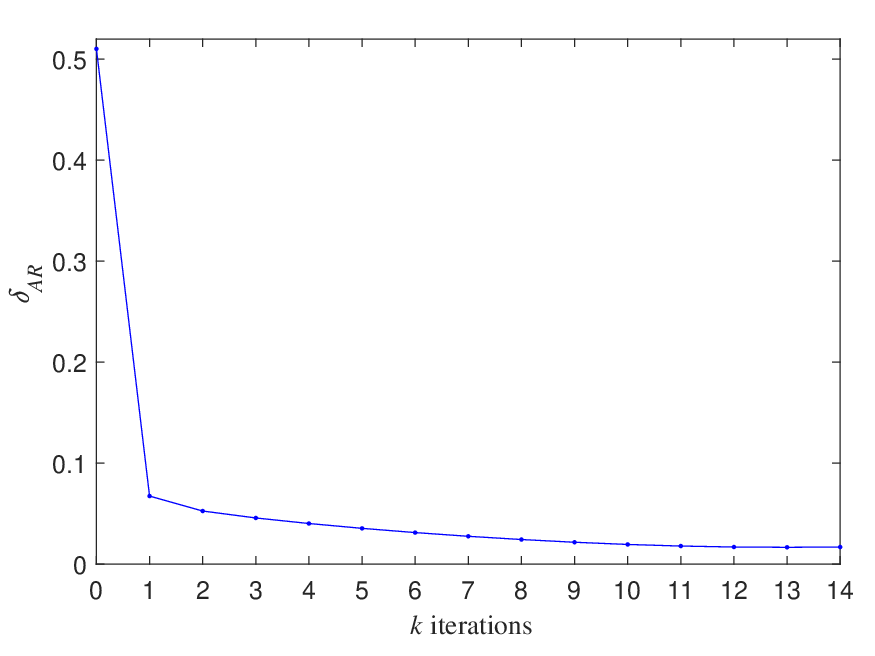}
  \caption{Estimation error $\delta_A$ against the iteration number $k$}
  \label{Fig_error_k}
\end{figure}

\begin{figure}[!hbt]
  \centering
  \includegraphics[width=0.8\hsize]{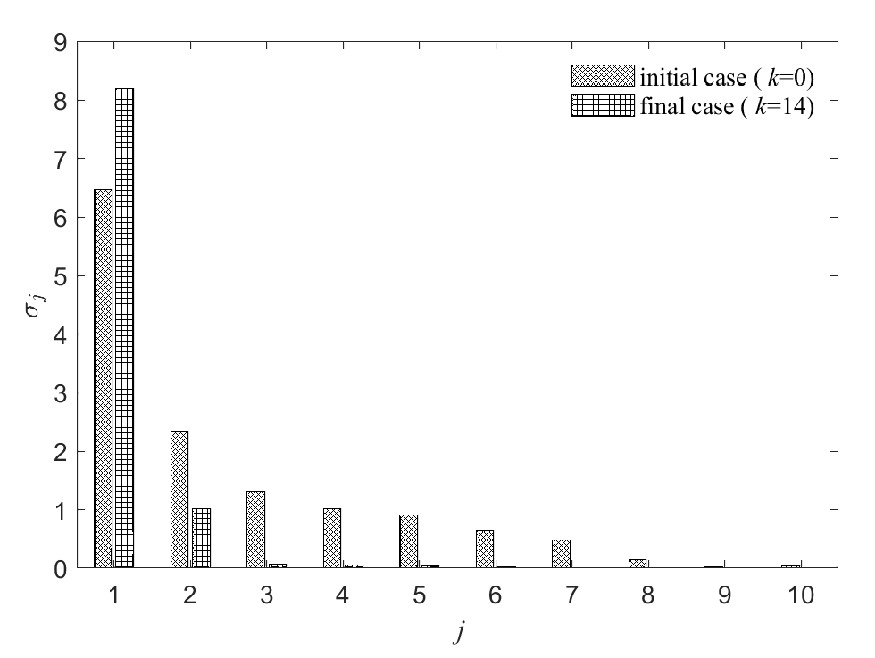}
  \caption{Singular values
of $\hat{\bfL}_{{\theta}_A}$}
  \label{Fig_sigluar_AKA}
\end{figure}

In the case of $\lambda=0.60$,
the parameter estimation errors $\delta_{AR}:=\|\bftheta_A-\hat{\bftheta}_A^k\|_F/\|\bftheta_A\|_F$ of the graphical AR process against the iteration number $k$ are shown in Fig. \ref{Fig_error_k},
where the initial value $\hat{\bftheta}_A^0$ is estimated
from (\ref{S_L2}) 
as described in Section~\ref{sec:ResultsAR}.
The estimation error decreases as the number of iterations increases.
After 14 iterations, a stable estimation
error is reached as $\delta_{AR}=1.69\%$,
and the corresponding AR parameter estimates including the positions and values of the non-zero entries are tabulated in Table~\ref{tab2}.
Let $\hat{\bfL}_{{\theta}_A}:={\hat{\bftheta}_A\hat{{\boldsymbol{K}}}\hat{\bftheta}_A^{\T}}-\bfI_n$ represent the estimate of the low-rank matrix involved in (\ref{phi_ast}) and (\ref{rankm1}).
Fig. \ref{Fig_sigluar_AKA} shows
the singular values $\sigma_j$ ($j=1,2,\cdots,10$)
of $\hat{\bfL}_{{\theta}_A}$
computed with the initial estimate $\hat{\bftheta}_A^0$
and final estimate $\hat{\bftheta}_A^{14}$, respectively.
The number of  non-zero singular values in the final case ($k=14$) is obviously reduced compared with that of the initial case ($k=0$),
indicating that a lower order model is obtained.
Combining the results shown in Fig. \ref{Fig_error_k}, Table~\ref{tab2} and Fig. \ref{Fig_sigluar_AKA}, we can see that the implementation of the reweighted trace
heuristic (\ref{bfthetaAk1})--(\ref{bfW2k1})
can effectively reduce the rank of the  primal low-rank solution estimated based on (\ref{S_L2}),
 following the improvement of the estimation accuracy.
 Finally, the parameter estimates with a higher accuracy can be achieved.
 \begin{table*}[!htb]
  \centering
  \caption{The estimated sparse parameter matrices of the graphical AR process}
  \label{tab2}
 \renewcommand{\arraystretch}{1}
\footnotesize \doublerulesep 0.25pt \tabcolsep 18pt 
  \begin{tabular}{ccccccc}\hline
  AR parameter matrices & $\hat{\bfA}_1$ & $\hat{\bfA}_1$& $\hat{\bfA}_1$&   $\hat{\bfA}_2$ & $\hat{\bfA}_2$& $\hat{\bfA}_2$ \\\hline
  Positions of non-zeros & (1,6)  & (9,4)     &  (10,3)    &  (2,5) & (5,7) & (8,1)         \\
   Estimated values &1.0324   & 0.9852   & 0.9776   & 1.0321     & 1.0243   & 0.9661      \\
   True values &1    & 1   & 1   & 1      & 1   & 1      \\\hline
  \end{tabular}
\end{table*}
\begin{figure}[!hbt]
  \centering
  \includegraphics[width=0.8\hsize]{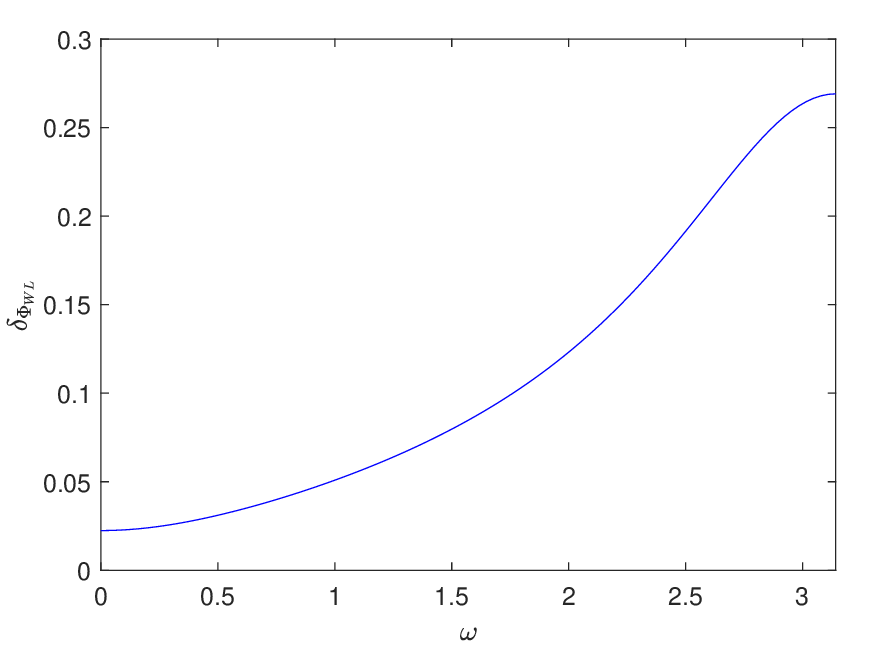}
  \caption{Estimation error $\delta_{\Phi_{WL}}(e^{{\rm{j}} \omega})$ versus $\omega \in[0,\pi]$}
  \label{Fig_error_omega}
\end{figure}

To measure the identification accuracy of the latent-variable parameters,
we compare the related estimates $\hat{\bfL}$ and $\hat{\bf\Phi}_{W_L}(e^{{\rm{j}} \omega}):=\Delta(e^{{\rm{j}} \omega})\hat{\bfL}\Delta^\ast(e^{{\rm{j}} \omega})$
with their true values.
The estimate $\hat{\bfL}$ has a unique non-zero eigenvalue 8.8456,
which is close to the true value 9.2960,
that is,
the largest eigenvalue of ${\bfL}$.
Since the rank of $\hat{\bfL}$ is consistent with the number of latent variables,
the exact number of latent variables is recovered as
$\hat{l}=1$.
This coincides with the analysis of the result in Table~\ref{tab1}.
The  spectrum estimation errors $\delta_{\Phi_{WL}}(e^{{\rm{j}} \omega}):=\|{\bf\Phi}_{W_L}(e^{{\rm{j}} \omega})-\hat{\bf\Phi}_{W_L}(e^{{\rm{j}} \omega})\|_F/\|{\bf\Phi}_{W_L}(e^{{\rm{j}} \omega})\|_F$
of the latent variables
at each frequency $\omega$ are plotted in Fig. \ref{Fig_error_omega},
showing that effective latent-variable parameter estimates can be obtained.

From this simulation,
we can conclude that for the \textcolor{blue}{graphical
AR Gaussian model with dynamical latent variables},
using the proposed sparse plus low-rank identification approach,
the graphical structure can be recovered,
and the dynamics of the AR process and latent variables
can be effectively estimated.

\textcolor{blue}{
\subsection{Simulation Example 2}
The data base used in this study consists of time series of daily stock market indices at closing time, in terms of local currency units, of fourteen financial markets.
The fourteen countries and their respective price indices are:
China (SSE Composite index denoted CH),
Hong Kong (Hang Seng index denoted HK),
Taiwan (TSEC weighted index denoted TA),
Japan (Nikkei 225 index denoted JA),
Korea (KOSPI Composite index denoted KO),
Belgium (BEL 20 index denoted BE),
Austria (ATX index denoted AS),
Germany (DAX index denoted GE),
France (CAC 40 index denoted FR),
Netherlands (AEX index denoted NL),
Unites States (S\&P500 denoted US),
Canada (S\&PTSX Composite index denoted CA),
Mexico (IPC index denoted ME) and Brazil (IBOVESPA index denoted BR).
The data are obtained from the website
at http://finance.yahoo.com/.
The sample period is from 4th
January 2018 up to 5th December 2019.
For each index, we compute the return
between trading day $t-1$ and $t$ as log differences:
$r(t)=100[\ln p(t)-\ln p(t-1)]$ with $p(t)$ closing price on day $t$ \cite{Abdelwahab2008_fm}.
Note that some missing index values are replaced
by the corresponding last trading day's values, and the return is zero.
The obtained data sequence has length $N=499$.}

\textcolor{blue}{
In this simulation example, the graphical AR model without latent variable $\sum_{j=0}^{2}{\boldsymbol{A}}_j{\boldsymbol{r}}(t-j)={\boldsymbol{w}}(t)$
and latent-variable graphical AR models 
$\sum_{j=0}^{2}{\boldsymbol{A}}_j{\boldsymbol{r}}(t-j)=\sum_{i=0}^{p_2}{\boldsymbol{W}}_{L,i}{\boldsymbol{x}}(t-i)+{\boldsymbol{w}}(t)$
with $p_2=0,1,2$
are adopted to model the time series consisting of all the fourteen daily financial return components
${\boldsymbol{r}}(t)=[r_{\rm{CH}}(t),r_{\rm{HK}}(t),\cdots,r_{\rm{BR}}(t)]\in \mathbb{R}^{14}$,
 and identify the interaction between financial markets by solving the regularized maximum likelihood method \cite{Songsiri2010_TSAR} and our proposed sparse plus low-rank identification algorithm, respectively.
The estimated models of different orders with the lowest fitness score are tabulated in Table~\ref{models_p}.
It can be seen that the introduction of latent variables can reduce the graphical complexity,  the number of identified interactions between financial markets is reduced from 19 to 5.
For estimated latent-variable graphical AR models,
some conditional dependences related to Hong Kong market cannot be characterized by the latent variable,
which accords with the empirical results analysed in \cite{Abdelwahab2008_fm}
such as the highest connection of the Japan/Hong Kong association.
The model fitness scores for graphical AR model without latent variables and latent-variable graphical AR model with $p_2=0$ are close. In addition,
along with the increase of $p_2$, the score is further reduced.
This implies that the dynamical structure of the latent variable can improve
data fitness.
\begin{table*}[!htb]
  \centering
  \caption{Estimated graphical models for the international financial stock returns data}
  \label{models_p}
 \renewcommand{\arraystretch}{1.3}
\footnotesize \doublerulesep 0.2pt \tabcolsep 2pt 
  \begin{tabular}{ccccc}\hline
  &\multirow{2}{*}{graphical AR model without latent variables}
  & \multicolumn{3}{c}{latent-variable graphical AR models}\\
  & &$p_2=0$  &$p_2=1$ &$p_2=2$ \\\hline
  $f$  &240.22 &239.85 &226.71 &214.55  \\
  $\vert\mathcal E_n\vert$ &19 &5&5&5\\
  $\mathcal E_n$ &\tabincell{c}{(CH,JA),\ (CH,BE),\ (HK,TA),\ (HK,JA),\ (HK,GE),\ (HK,US),\ (HK,ME),\\ (HK,BR),\ (TA,JA),\ (TA,US),\ (TA,BR),\ (JA,KO),\ (JA,GE),\\
  (JA,US),\ (JA,BR),\ (KO,US),\ (GE,US), \ (GE,BR),\ (US,BR)}
  & &
  \tabincell{c}{(HK,JA),\ (HK,GE),\ (HK,US),\\ (HK,BR),\ (TA,US)}
  \\\hline
  \end{tabular}
\end{table*}}

\section{Conclusions}
\label{sec:Conclusions}
This paper has developed an identification approach for \textcolor{blue}{dynamical latent-variable graphical AR Gaussian
models}.
A new sparse plus low-rank optimization paradigm
has been established by \textcolor{blue}{combining} the sparse part of the inverse spectrum corresponding to the observable variables
with the low-rank spectrum representation of the
latent variables.
By employing the reweighted trace heuristic,
the graphical structure among the observable variables as well as the sparse AR parameters have been obtained.
Then based on the identified AR graphical part,
the dynamical model of latent variables including the dimension and parameters have been estimated by minimizing the rank of the latent-variable spectrum.
Simulation examples have been given to
demonstrate the effectiveness of the proposed approach.

The proposed identification procedure is suboptimal but can achieve good identification performance according to the simulation results.
When extending AR models to ARMA models,
the corresponding graphical ARMA models with dynamic latent variables have more complicated sparse plus low-rank structures and the additional MA part makes the low-rank structure of the dynamic latent variables difficult to be estimated separately.
Therefore, this
challenging identification problem will be investigated in our future work.

\textcolor{blue}{
\section*{Appendix}\label{secA1}
\subsection*{A. Proof of Proposition 4}
Since $\bfQ=\mathcal{D}(\bfX)$,
Problem (\ref{S_L1}) can be rewritten as
\begin{eqnarray}
\label{S_LQ}
\begin{split}
&\mathop{\rm{min}}\limits_{\bfX,\bfQ}\ {\rm{tr}}({\boldsymbol{K}}\bfX)-n+
\gamma h_\infty(\bfQ),\\
&{\rm{s.t.}}\ \ \ \bfX \geq 0,\ \ \bfX_{0,0}=\bfI_n,\ \  \bfQ=\mathcal{D}(\bfX).
\end{split}
\end{eqnarray}
By introducing Lagrange multipliers $\bfZ\in \mathbb{M}^{n,p_1}$, $\bfU\in\mathbb{S}^{n(p_1+1)}$ and $\bfP\in\mathbb{S}^{n}$, the
Lagrangian function is
\begin{eqnarray}
&&{\mathcal L}(\bfX,\bfQ,\bfZ,\bfU,\bfP)\non\\
&=&{\rm{tr}}({\boldsymbol{K}}\bfX)-n+
\gamma h_\infty(\bfQ)-\langle\bfU,\bfX\rangle\non\\
&&+\langle\bfP,\bfI_n-\bfX_{0,0}\rangle+\langle\bfZ,\mathcal{D}(\bfX)-\bfQ\rangle\non\\
&=&\langle\bfK+\mathcal{T}(\bfZ)-\bfU,\bfX\rangle-\langle\bfP,\bfX_{0,0}\rangle\non\\
&&+\gamma h_\infty(\bfQ)-\langle\bfZ,\bfQ\rangle+{\rm{tr}}(\bfP)-n,\non
\end{eqnarray}
where we exploited the fact that the mappings $\mathcal{T}$ and $\mathcal{D}$ are adjoints, that is, $\langle\bfZ,\mathcal{D}(\bfX)\rangle=\langle\mathcal{T}(\bfZ),\bfX\rangle$.
The dual function is the infimum of ${\mathcal L}$ over $\bfX$ and $\bfQ$.
As shown in \cite{Songsiri2010_TSAR}, the minimization over
$\bfQ$ is bounded below only if
\begin{eqnarray}
\label{Qcon1}
&&{\rm{diag}}(\bfZ_j)=0,\ \ j=0,\cdots,p_1,\\
\label{Qcon2}
&&{\sum_{j=0}^{p_1}}(\vert[\bfZ_j]_{kq}\vert+\vert[\bfZ_j]_{qk}\vert)\leq\gamma,
\ \ k\neq q,
\end{eqnarray}
and the corresponding infimum is zero.
The partial minimization of ${\mathcal L}$ over $\bfQ$ is thereby given by
\begin{eqnarray}
\mathop{\rm{inf}}\limits_{\bfQ}{\mathcal L}=
\left\{
\begin{aligned} &\langle\bfK+\mathcal{T}(\bfZ)-\bfU,\bfX\rangle-\langle\bfP,\bfX_{0,0}\rangle+{\rm{tr}}(\bfP)-n,\\ &\ \ \ \ \ \ \ \ \ \ \ \ \ \ \ \ \ \ \ \ \ \ \ \ \ \ \ \ \ \ \ \ \ \ \ \ \ \ \ \ \ \ \ \ \ \ \ \ \ \ \ \ \  (\ref{Qcon1}),(\ref{Qcon2}) \\
&-\infty,\ \ {\rm{otherwise}}.\non
\end{aligned}
\right.
\end{eqnarray}
The minimization  of ${\mathcal L}$ over $\bfX$ is bounded below only if
\begin{eqnarray}
\label{X00con}
&&(\bfK+\mathcal{T}(\bfZ)-\bfU)_{0,0}-\bfP=0,\\
\label{Xcon}
&&(\bfK+\mathcal{T}(\bfZ)-\bfU)_{i,j}=0,\ i\neq 0,\ j\neq 0,
\end{eqnarray}
and the corresponding infimum is zero. Thus, the dual function is
\begin{eqnarray}
\mathop{\rm{inf}}\limits_{\bfX,\bfQ}\ {\mathcal L}=
\left\{
\begin{aligned} &{\rm{tr}}(\bfP)-n,\ \ (\ref{Qcon1})-(\ref{Xcon}) \\
&-\infty,\ \ {\rm{otherwise}}.\non
\end{aligned}
\right.
\end{eqnarray}
The dual problem is to maximize the dual function over $\bfZ$, $\bfU$ and $\bfP$ subject to $\bfU \geq 0$. By eliminating the slack variable $\bfU$, the dual problem can be expressed as
\begin{eqnarray}
\label{dual}
\begin{split}
&\mathop{\rm{max}}\limits_{{\bfP},{\bfZ}}\ {\rm{tr}}(\bfP)-n,\\
&{\rm{s.t.}}\ \ \left[\begin{array}{cc}
           \bfP     &  {\bf{0}}_{n\times np_1}        \\
             {\bf{0}}_{np_1\times n}       &  {\bf{0}}_{np_1\times np_1}  \end{array}\right]\leq \bfK+\mathcal{T}(\bfZ),\\
&\ \ \ \  \ \ {\rm{diag}}(\bfZ_j)=0,\ \ j=0,\cdots,p_1,\\
&\ \ \ \ \ \ {\sum_{j=0}^{p_1}}(\vert[\bfZ_j]_{kq}\vert+\vert[\bfZ_j]_{qk}\vert)\leq\gamma,
\ \ k\neq q.
\end{split}
\end{eqnarray}
The primal problem (\ref{S_LQ}) is strictly feasible (pick $\bfX=\bfI_{n(p_1+1)}$),
thus Slater's condition holds.
Accordingly, the
duality gap between the primal and dual problems is equal to zero,
that is
\begin{eqnarray}
\label{Zerodualitygap1}
\langle\bfU,\bfX\rangle=
\left\langle\bfK+\mathcal{T} (\bfZ)-\left[\begin{array}{cc}
           \bfP     &  {\bf{0}}_{n\times np_1}        \\
             {\bf{0}}_{np_1\times n}       &  {\bf{0}}_{np_1\times np_1}  \end{array}\right],\bfX\right\rangle=0.
\end{eqnarray}
Since $\bfU \geq 0$, $\bfX \geq 0$, (\ref{Zerodualitygap1})
is equivalent to
\begin{eqnarray}
\label{Zerodualitygap2}
\bfU\bfX=
\left(\bfK+\mathcal{T} (\bfZ)-\left[\begin{array}{cc}
           \bfP     &  {\bf{0}}_{n\times np_1}        \\
             {\bf{0}}_{np_1\times n}       &  {\bf{0}}_{np_1\times np_1}  \end{array}\right]\right)\bfX=0.
\end{eqnarray}
When $p_1=0$, the zero duality gap (\ref{Zerodualitygap2}) is
\begin{eqnarray}
\bfU_{0,0}\bfX_{0,0}=
(\bfR_0+\bfZ_0-\bfP)\bfX_{0,0}=0,\non
\end{eqnarray}
and the dual problem (\ref{dual}) reduces to
\begin{eqnarray}
\label{dualp0}
\begin{split}
&\mathop{\rm{max}}\limits_{{\bfP},{\bfZ}}\ {\rm{tr}}(\bfP)-n,\\
&{\rm{s.t.}}\
           \bfP     \leq \bfR_0+\bfZ_0,\  {\rm{diag}}(\bfZ_0)=0,\ \vert[\bfZ_0]_{kq}\vert\leq\gamma/2,\  k\neq q.
\end{split}
\end{eqnarray}
In this case, the optimal solution of the primal problem (\ref{S_LQ}) is $\bfX=\bfX_{0,0}=\bfI_n$,
and the corresponding optimal value of the objective function
is ${\rm{tr}}(\bfR_0)-n$ ($\bfQ=\bfX_{0,0}=\bfI_n$, $h_\infty(\bfQ)=0$).
Since the optimal value coincides with that of the dual problem (\ref{dualp0}),
we can obtain that
\begin{eqnarray}
\bfP=\bfR_0,\ \ \bfZ_0=0,\ \ \bfU_{0,0}=0.\non
\end{eqnarray}
Then the dual problem (\ref{dual}) can be rewritten as
\begin{eqnarray}
\label{dual2}
\begin{split}
&\mathop{\rm{max}}\limits_{{\bfZ}}\ {\rm{tr}}(\bfR_0)-n,\\
&{\rm{s.t.}}\ \ \left[\begin{array}{cc}
           \bfR_0     &  {\bf{0}}_{n\times np_1}        \\
             {\bf{0}}_{np_1\times n}       &  {\bf{0}}_{np_1\times np_1}  \end{array}\right]\leq \bfK+\mathcal{T}(\bfZ),\\
&\ \ \ \  \ \ {\rm{diag}}(\bfZ_j)=0,\ \ j=0,\cdots,p_1,\\
&\ \ \ \ \ \ {\sum_{j=0}^{p_1}}(\vert[\bfZ_j]_{kq}\vert+\vert[\bfZ_j]_{qk}\vert)\leq\gamma,
\ \ k\neq q,\non
\end{split}
\end{eqnarray}
which is strictly feasible if $\bfK>0$ (take $\bfZ=0$).
The solutions $\bfX^\circ$, $\bfQ^\circ$, and $\bfZ^\circ$ of the primal and dual problems are characterized by the following set of necessary and sufficient optimality (or Karush-Kuhn-Tucker) conditions.\\
{\bf{Primal feasibility}}
\begin{eqnarray}
\bfX^\circ \geq 0,\ \bfX^\circ_{0,0}=\bfI_n,\ \bfQ^\circ=\mathcal{D}(\bfX^\circ).\non
\end{eqnarray}
{\bf{Dual feasibility}}
\begin{eqnarray}
&&\left[\begin{array}{cc}
           \bfR_0    &  {\bf{0}}_{n\times np_1}        \\
             {\bf{0}}_{np_1\times n}       &  {\bf{0}}_{np_1\times np_1}  \end{array}\right]\leq \bfK+\mathcal{T}(\bfZ^\circ),\non\\
          &&{\rm{diag}}(\bfZ^\circ_j)=0,\ \ j=0,\cdots,p_1,\non\\ &&{\sum_{j=0}^{p_1}}(\vert[\bfZ^\circ_j]_{kq}\vert+\vert[\bfZ^\circ_j]_{qk}\vert)\leq\gamma,
\ \ k\neq q.\non
\end{eqnarray}
{\bf{Zero duality gap}}
\begin{eqnarray}
\label{Zerodualitygap4}
\left(\bfK+\mathcal{T} (\bfZ^\circ)-\left[\begin{array}{cc}
           \bfR_0     &  {\bf{0}}_{n\times np_1}        \\
             {\bf{0}}_{np_1\times n}       &  {\bf{0}}_{np_1\times np_1}  \end{array}\right]\right)\bfX^\circ=0.
\end{eqnarray}
\begin{Lemma} \cite{Songsiri2010_gmar}
 If $\mathcal{T}(\bfS)$ is a symmetric block-Toeplitz matrix, with $\bfS\in \mathbb{M}^{n,p}$, and
\begin{eqnarray}
&&\mathcal{T}(\bfS)\geq \left[\begin{array}{cc}
           \bfW     &  {\bf{0}}_{np\times np}        \\
             {\bf{0}}_{np\times np}       &  {\bf{0}}_{np\times np}  \end{array}\right],\non
\end{eqnarray}
for some $\bfW\in\mathbb{S}^{n}$ positive definite, then $\mathcal{T}(\bfS)>0$.
\end{Lemma}
If $\bfR_0>0$ (this can be follows from $\bfK>0$), by Lemma 5, we have that $\bfK+\mathcal{T} (\bfZ^\circ)>0$.
Accordingly, the rank of
\begin{eqnarray}
\bfK+\mathcal{T} (\bfZ^\circ)-\left[\begin{array}{cc}
           \bfR_0     &  {\bf{0}}_{n\times np_1}        \\
             {\bf{0}}_{np_1\times n}       &  {\bf{0}}_{np_1\times np_1}  \end{array}\right]\non
\end{eqnarray}
is at least $np_1$.
The complementary slackness condition (\ref{Zerodualitygap4}) implies that $\bfX^\circ$ has rank at most equal to $n$.
On the other hand, it is obviously that ${\rm{rank}}(\bfX^\circ)\geq n$ since $\bfX^\circ_{0,0}=\bfI_n$.
Therefore, it can be concluded that ${\rm{rank}}(\bfX^\circ)=n$.
There exists $\hat{\bftheta}_A:=[{\bfI}_n,\hat{\bfA}_1,\cdots,\hat{\bfA}_{p_1}]\in\mathbb{R}^{n\times n(p_1+1)}$ full
row rank such that $\bfX^\circ=\hat{\bftheta}_A^{\T}\hat{\bftheta}_A$,
and (\ref{Zerodualitygap4}) implies
\begin{eqnarray}
\label{Zerodualitygap_B}
[\bfK+\mathcal{T}(\bfZ^\circ)]\hat{\bftheta}_A^{\T}=\left[\begin{array}{c}
           \bfR_0             \\
             {\bf{0}}_{np_1\times n} \end{array}\right].
\end{eqnarray}
Since $\bfK+\mathcal{T}(\bfZ^\circ)> 0$,
Equation (\ref{Zerodualitygap_B}) admits a
unique solution $\hat{\bftheta}_A$.
The uniqueness of $\bfX^\circ$ follows from the uniqueness of $\hat{\bftheta}_A$.
\ \ \ \ \ \ \ \ \ \ \ \ \ \ \ \ \ \ \ \ \ \ \ \ \ \ \ \ \ \ \ \ \ \ \ \ \ \ \ \ \ \ \ \ \ \ \ \ \ \ \ \ \ \ \ \ \ \ \
$\blacksquare$}

\textcolor{blue}{
\subsection*{B. Demonstration of Theorem 6}
Two assumptions are introduced in \cite{Cand2009_MC}
about an $n_1\times n_2$, rank $r$ matrix $\bfM$ whose SVD is given by
\begin{eqnarray}
\bfM&=&\bfU_{\bfM}\Sigma_{\bfM}\bfV_{\bfM}^{\T}=\sum_{k=1}^{r}\sigma_{\bfM,k}\bfu_{\bfM,k}\bfv_{\bfM,k}^{\T}\non
\end{eqnarray}
and with column and row spaces denoted by $U_{\bfM}$ and $V_{\bfM}$,
respectively.\\
${\bfA0}$ The coherences obey ${\rm{max}}(\mu(U_{\bfM}),\mu(V_{\bfM}))\leq \mu_0$ for some
$\mu_0>0$.\\
${\bfA1}$ The $n_1\times n_2$ matrix $\sum_{1\leq k\leq r}\bfu_{\bfM,k}\bfv_{\bfM,k}^{\T}$ has a maximum entry bounded by $\mu_1\sqrt{r/(n_1n_2)}$  in absolute value for some $\mu_1>0$.
\begin{Theorem}\cite{Cand2009_MC}
\label{Theorem2}
Let $\bfM$ obey assumptions ${\bfA0}$ and ${\bfA1}$
and put $\tilde{n}={\rm{max}}(n_1,n_2)$.
 Suppose we observe $m$ entries of $\bfM$ with locations sampled uniformly at random and let $\Omega$ be a set composed of locations corresponding to the observed entries. Then there exist constants $C$, $c$ such that if
\begin{eqnarray}
m\geq C{\rm{max}}(\mu_1^2,\mu_0^{1/2}\mu_1,\mu_0n^{1/4})\tilde{n}r(\beta \log{\tilde{n}})\non
\end{eqnarray}
for some $\beta>2$, then the minimizer to the problem
\begin{eqnarray}
 \begin{split}
& \mathop{\rm{min}}\limits_{\bfT}\ \|\bfT\|_\ast,\\
&{\rm{s.t.}} \ \ {\bfT}_{ij}={\bfM}_{ij},\ \ (i,j)\in \Omega\non
\end{split}
\end{eqnarray}
 is unique
and equal to $\bfM$ with probability at least $1-c\tilde{n}^{-\beta}$.
For $r\leq \mu_0^{-1}\tilde{n}^{1/5}$
this estimate can be improved to
\begin{eqnarray}
m\geq C\mu_0\tilde{n}^{6/5}r(\beta \log{\tilde{n}})\non
\end{eqnarray}
with the same probability of success.
\end{Theorem}
In our problem setting (\ref{rankm2}),
the variable $\bfX_L(\bftheta_A)$ to be solved is a square matrix of dimension $\tilde{n}=n_1=n_2=n(p_1+2)$ and rank $r=l(p_2+1)+n(p_1+1)$ if $l(p_2+1)<n$.
We have observed subblocks of $\bfX_L(\bftheta_A)$:
the top left subblock $\bfK^{-1}\in\mathbb{S}^{n(p_1+1)}$,
the bottom right subblock $\bfI_n\in\mathbb{S}^{n}$
and the first block of $\bftheta_A(:,1:n)=\bftheta_A^{\T}(1:n,:)=\bfI_n\in\mathbb{S}^{n}$ (provided in the constraint of (\ref{rankm2})).
Suppose we have obtained the graph topology estimate ${\rm{supp}}(\hat{{\bf\Phi}}_S)$,
that is the locations of zero entries of $\bftheta_A$ except its first block are observed (also provided in the constraints),
then $m=\tilde{n}^2-2p_1\vert{\rm{supp}}(\hat{{\bf\Phi}}_S)\vert$ entries of $\bfX_L(\bftheta_A)$
are known.
When ${\rm{supp}}(\hat{{\bf\Phi}}_S)$ is close to or equal to the sparsity pattern of true ${\bf\Phi}_S$ in (\ref{Phi_S}),
$m$ will be large and
the observed entries are close to uniformly distributed with their values close to or equal to the corresponding true values.
In analogy with $\bfM$,
if $\bfX_L(\bftheta_A)$ obey assumptions ${\bfA0}$ and ${\bfA1}$, and its rank $r$, dimension $\tilde{n}$ and number of observed entries $m$ satisfy the quantitative relation in Theorem \ref{Theorem2},
Theorem \ref{Theorem1} is established.
\ \ \ \ \ \ \ \ \ \ \ \ \ \ \ \ \ \ \ \ \ \ \ \ \ \ \ \ \ \ \ \ \ \ \ \ \ \ \ \ \ \
\ \ \ \ \ \ \ \ \ \ \ \ \ \ \ \ \ \ \ \ \ \ \
$\blacksquare$}

\bibliographystyle{plain}

\end{document}